\documentclass[a4paper,11pt]{article}
\usepackage{dcolumn}
\usepackage{bm}
\usepackage{booktabs}
\usepackage{graphics,subfigure,axodraw}
\usepackage{graphicx,psfrag}
\usepackage{epsfig}
\usepackage{color}
\usepackage{amsmath,amssymb}
\usepackage{geometry}
\usepackage{slashed}
\usepackage{cite}
\usepackage{mathrsfs}

\allowdisplaybreaks[3]

\newcommand{\newc}{\newcommand}
\newc{\ie}{\ensuremath{\mathrm{i}}}
\newc{\eu}{\ensuremath{\mathrm{e}}}
\newc{\cw}[1][{}]{\ensuremath{c^{#1}_\mathrm{w}}}
\newc{\sw}[1][{}]{\ensuremath{s^{#1}_\mathrm{w}}}
\newc{\tw}[1][{}]{\ensuremath{\tan^{#1} \theta_{\mathrm{w}}}}
\newc{\ctw}[1][{}]{\ensuremath{\cot^{#1}\theta_{\mathrm{w}}}}
\newc{\real}{\ensuremath{\mathfrak{Re}}}
\newc{\imag}{\ensuremath{\mathfrak{Im}}}
\newc{\cv}[1][{}]{\ensuremath{\cos^{#1} \beta}}
\newc{\sv}[1][{}]{\ensuremath{\sin^{#1} \beta}}
\newc{\tv}[1][{}]{\ensuremath{\tan^{#1} \beta}}
\newc{\M}{\ensuremath{\mathcal{M}}}
\newc{\ssl}[1]{\ensuremath{\slashed{#1}}}
\newc{\half}{\ensuremath{\frac{1}{2}}}
\newc{\photino}{\widetilde{\gamma}}
\newc{\bino}{\widetilde{\cal B}}
\newc{\wino}{\widetilde{\cal W}}
\newc{\gluino}{\widetilde{\cal G}}
\newc{\rpv}{{\mbox{${\not\!\!R_p}$}}}
\newc{\cL}{{\cal L}}
\newc{\x}[1]{\ensuremath{\tilde{\chi}^0_{#1}}}
\newc{\GeV}{\ensuremath{\,\mathrm{GeV}}}
\newc{\fb}{\ensuremath{\,\mathrm{fb}}}
\newc{\pb}{\ensuremath{\,\mathrm{pb}}}
\newc{\erg}{\ensuremath{\,\mathrm{erg}}}
\DeclareMathOperator*{\diag}{diag}
\def\lsim{\raise0.3ex\hbox{$\;<$\kern-0.75em\raise-1.1ex\hbox{$\sim\;$}}}
\def\gsim{\raise0.3ex\hbox{$\;>$\kern-0.75em\raise-1.1ex\hbox{$\sim\;$}}}
\newc{\MeV}{\,{\mathrm{MeV}}}
\newc{\lsp}{{{\tilde\chi}}}
\newc{\lam}{\lambda}
\newc{\ra}{\rightarrow}
\newc{\htext}[1]{{\color{red}  #1}}
\newc{\htextb}[1]{{\color{blue}  #1}}
\newc{\sweff}{\sin^2\theta_{\mathrm{eff}}}
\newc{\cha}[1]{\tilde \chi^\pm_{#1}}
\newc{\mcha}[1]{m_{\tilde \chi^\pm_{#1}}}
\newc{\neu}[1]{\tilde \chi^0_{#1}}
\newc{\mneu}[1]{m_{\tilde \chi^0_{#1}}}
\def\refeq#1{\mbox{Eq.~(\ref{#1})}}
\def\refeqs#1{\mbox{Eqs.~(\ref{#1})}}
\def\reffi#1{\mbox{Fig.~\ref{#1}}}

\def\refta#1{\mbox{Table~\ref{#1}}}

\def\refse#1{\mbox{Sect.~\ref{#1}}}

\def\refsubse#1{\mbox{Subsect.~\ref{#1}}}

\def\citere#1{\mbox{Ref.~\cite{#1}}}
\def\citeres#1{\mbox{Refs.~\cite{#1}}}

\newcommand{\MZ}{M_Z}
\newcommand{\MW}{M_W}

\graphicspath{{figs/}}

\begin{document}

\thispagestyle{empty}

\def\thefootnote{\fnsymbol{footnote}}

\begin{flushright}
Bonn-TH-2008-11 \\
DCPT/09/04 \\
DESY 08-190\\
IPPP/09/02 \\
arXiv:0901.3485 [hep-ph]
\end{flushright}

\vspace{1cm}

\begin{center}

{\Large\sc {\bf Mass Bounds on a Very Light Neutralino}}

\vspace{1cm}

{\sc
Herbi~K.~Dreiner$^{1}$%
\footnote{email: dreiner@th.physik.uni-bonn.de}%
, Sven~Heinemeyer$^{2}$%
\footnote{email: Sven.Heinemeyer@cern.ch}%
, Olaf~Kittel$^{3}$%
\footnote{email: kittel@th.physik.uni-bonn.de}%
, Ulrich~Langenfeld$^{4}$%
\footnote{email: Ulrich.Langenfeld@desy.de}%
,\\[.5em] Arne~M.~Weber$^{5}$%
\footnote{email: arne@mppmu.mpg.de}%
~and Georg Weiglein$^{6}$%
\footnote{email: Georg.Weiglein@durham.ac.uk}
}

\vspace*{.7cm}
\date{\today}
{\sl
$^1$Bethe Center for Theoretical Physics \& Physikalisches Institut der Universit\"at Bonn, Germany

\vspace*{0.1cm}

$^2$Instituto de F\'isica de Cantabria (CSIC-UC), Santander,  Spain

\vspace*{0.1cm}

$^3$Departamento de F\'isica Te\'orica y del Cosmos and CAFPE, \\
Universidad de Granada, E-18071 Granada, Spain

\vspace*{0.1cm}

$^4$DESY Zeuthen, Zeuthen, Germany

\vspace*{0.1cm}

$^5$Max-Planck-Institut f\"ur Physik (Werner-Heisenberg-Institut),
F\"ohringer Ring 6, \\
D--80805 M\"unchen, Germany%
\footnote{former address}

\vspace*{0.1cm}

$^6$IPPP, University of Durham, Durham DH1~3LE, UK
}

\end{center}

\vspace*{0.1cm}

\begin{abstract}
\noindent
Within the Minimal Supersymmetric Standard Model (MSSM) we
systematically investigate the bounds on the mass of the lightest
neutralino. We allow for non-universal gaugino masses and thus even
consider massless neutralinos, while assuming in general that
$R$-parity is conserved. Our main focus is on laboratory constraints. We
consider collider data, precision observables, and also rare meson
decays to very light neutralinos. We then discuss the astrophysical
and cosmological implications. We find that a massless neutralino is
allowed by all existing experimental data and astrophysical and
cosmological observations.
\end{abstract}

\def\thefootnote{\arabic{footnote}}
\setcounter{page}{0}
\setcounter{footnote}{0}

\newpage

\section{Introduction}

{
\input paperdef
\subsection{Motivation}

The LHC is scheduled to start taking data in 2009, experimentally
opening the window to the TeV energy scale~\cite{lhc}. One of the main
goals is to search for physics beyond the standard model (SM). A
promising candidate is weak-scale supersymmetry (SUSY), which among
other attractive features stabilizes the hierarchy between the weak
scale and the Planck scale~\cite{Ellis:2007vz}.  Supersymmetry
predicts a large number of new particles, which if kinematically
accessible, should be observable at the
LHC~\cite{Haber:1984rc,Nilles:1983ge,Drees:2004jm,Baer:2006rs}.

The lightest supersymmetric particle, the LSP, plays a special role in
the search for supersymmetry at colliders. If weak-scale supersymmetry
is realised in nature, the production rate of the heavier squarks and
gluinos will be dominant, due to their strong interactions. But these
heavier particles rapidly cascade decay to the LSP in the detector,
typically with no detached vertex. Since this occurs in nearly all
supersymmetric events, the nature of the LSP and its behaviour are
decisive for all supersymmetric signatures at the LHC. Several LSP
candidates have been discussed in the literature,
\textit{e.g.} the lightest neutralino, the
gluino~\cite{Raby:1997bpa,Baer:1998pg}, the lightest
stau~\cite{Akeroyd:1997iq,Hirsch:2003fe,deGouvea:1998yp}, and the
lightest sneutrino~\cite{Chou:2000cy,deGouvea:2006wd}. In the case of
conserved proton hexality, P$_6$~\cite{Dreiner:2005rd,Dreiner:2007vp}
(or conserved $R$-parity\footnote{This is equivalent for the
renormalisable superpotential.} \cite{Farrar:1978xj}) the LSP (and the
proton) is stable. A stable light gluino LSP has been excluded by the
LEP data~\cite{light-gluino}. A stable (left-handed) sneutrino LSP is
also experimentally excluded~\cite{Hebbeker:1999pi}. Furthermore
cosmologically, a stable LSP must be electrically and colour
neutral~\cite{Ellis:1983ew}. This leaves as the most widely studied
candidate the lightest neutralino: $\neu{1}$, which is a very
promising dark matter
candidate~\cite{Pagels:1981ke,Goldberg:1983nd,Ellis:1983ew}.  Further
possibilities beyond the MSSM are the gravitino~\cite{gravitino} or
the axino~\cite{Rajagopal:1990yx}, which we do not discuss here. We
mention as an aside that in the case of ensuring the proton stability
via baryon-triality~\cite{Ibanez:1991pr,Dreiner:2006xw} the LSP
decays, and in principle any supersymmetric particle can be the LSP.
When embedding such models in minimal supergravity the stau, the
right-handed smuon, the right-handed selectron, the sneutrino, the
lightest sbottom or stop, as well as the right-handed sstrange or
sdown have been found to be consistent with
experiments~\cite{b3msugra,Dreiner:2008ca}.

\medskip

In this paper, we focus on a particular aspect of the neutralino LSP,
$\neu{1}$, namely `How light can it be?'\!. Our main interest hereby
are the laboratory constraints, \textit{i.e.}  `What can one learn
about the neutralino mass from \textit{collider} or
\textit{fixed-target experiments}?'. Thus we initially put aside all
aspects of the neutralino as a potential dark matter candidate. We
shall discuss the cosmological implications of a very light
neutralino towards the end of this paper. One should keep in mind
that potential cosmological constraints or requirements can be avoided
by adding a small (or large) amount of $R$-parity violation. Potential
other dark matter candidates in this case have been discussed in the
literature, see for example Refs.~\cite{Chun:1999cq,Lee:2007fw,Lee:2007qx}.
Unless otherwise indicated, we assume in our analyses that $R$-parity 
is conserved.

The particle data group (PDG) cites as the laboratory bound on the
lightest neutralino mass~\cite{Amsler:2008zz}
\begin{equation}
\mneu{1}>46\GeV    
\label{lep-bound}
\end{equation}
at 95\% C.L., which is based on the searches at LEP (the limit quoted
above was obtained in the analysis of the DELPHI
collaboration~\cite{Abdallah:2003xe}). This has been obtained by
taking into account the LEP searches for charginos, which yield lower
limits on the SU(2) gaugino mass parameter, $M_2$, and the higgsino
mass parameter, $\mu$.  Furthermore, this bound assumes an underlying
supersymmetric grand unified theory, based on a simple Lie group. In
this case, the U(1)$_Y$ gaugino mass parameter $M_1$ is related to
$M_2$ by
\begin{eqnarray}
\label{assump}
M_1 = \frac{5}{3}\tw[2] M_2 \approx \half M_2\,,
\end{eqnarray}
so that the experimental bound on $M_2$ implies a lower bound on
$M_1$.  Thus the chargino searches yield lower limits on $M_2$ and
$\mu$ and indirectly on $M_1$, which together give rise to a lower
bound on the mass of the lightest neutralino, see
Eq.~(\ref{lep-bound}), via the neutralino mass matrix, see
Eq.~(\ref{neutralinomatrix}) below.

It is the purpose of this paper to investigate the consequences of
dropping the assumption Eq.~(\ref{assump}), which invalidates the
bound in Eq.~(\ref{lep-bound}). Such a scenario could occur, for
instance, in the case of an underlying string unification with a
semi-simple gauged Lie group~\cite{strings}. In this more general
scenario, $M_1$ and $M_2$ are both free parameters.  As we shall see,
this additional freedom allows the LEP bound, Eq.~(\ref{lep-bound}),
to be avoided. This raises the question of the corresponding new lower
mass bound on the neutralino.  Such models have been considered in
Refs.~\cite{Hooper:2002nq,Belanger:2002nr,Bottino:2002ry,Belanger:2003wb,Bottino:2003iu,Bottino:2003cz,Lee:2007ai,Profumo:2008yg},
however while still requiring the neutralino to provide the entire
dark matter of the universe. As stated above, our main focus here are
the laboratory bounds. We systematically demonstrate that a
\textit{massless} neutralino is consistent with all present laboratory
data.  We shall consider bounds from direct collider searches,
precision electroweak data, and rare meson decays. We then consider
the astrophysical (supernova) and cosmological implications of a light
neutralino, with a particular focus on a massless neutralino.

\subsection{Outline and Connection to Previous Work}

In early supersymmetric model building, the superpartner of the
photon, the photino, $\tilde\gamma$, was often considered to be
massless or very light (mass below 1 keV)
\cite{Salam:1974xa,Fayet:1974pd}. The vanishing mass was guaranteed by
a global $R$-symmetry~\cite{Fayet:1974pd}, under which the parameters
of the supersymmetry transformation, the Grassman variables $\theta,\,
\bar\theta$, transform non-trivially. However, such an $R$-symmetry
has several problems. First, it prohibits all gaugino masses,
including the gluino mass. As mentioned above, a light gluino has been
excluded by LEP data~\cite{light-gluino}. Second, spontaneous breaking
of the global $R$-symmetry leads to a problematic axion. The axion and
the light gluino can only be evaded by large explicit breaking terms
which however render the original symmetry
meaningless~\cite{Chamseddine:1995gb}.  It is possible to gauge a U(1)
$R$-symmetry in N=1 local supersymmetry~\cite{Freedman:1976uk}.
However in order to cancel the mixed triangle anomalies, we must
extend the field content by non-singlet fields under the SM gauge
symmetry, or consider a family dependent
U$(1)_R$~\cite{Chamseddine:1995gb}, see also
Ref.~\cite{Barbieri:1984aj}.  In contrast, we shall here consider a
very light or massless neutralino, where the small mass comes about
through a modest amount of fine-tuning between the parameters $M_1$
and $M_2$ and consider no further symmetry. This is discussed in
detail in Sect.~\ref{framework}. It is then a question of
phenomenology, which we will discuss in detail in this paper, to see
whether such a light neutralino is consistent with all laboratory,
astrophysical and cosmological data. We would also like to
point out that moels with a light neutralino have recently been
obtained in the context of gauge mediated supersymmetry breaking,
where heavy messenger masses are provided by the adjoint Higgs field
of an underlying $SU(5)$ grand unified theory~\cite{Dudas:2008eq}.
For related work on light neutralinos in other extensions of
the MSSM see, for example, Refs.~\cite{Barger:2005hb,Gunion:2005rw}.

\medskip

Following up on the early supersymmetric model building, the
phenomenological implications of a stable light photino were mainly
discussed in the context of cosmology, namely as a dark matter
candidate\footnote{See also the later
work~\cite{Chung:1997rq,Kolb:1996jm,Farrar:1996jm}, where photinos of
mass $\mathcal{O}(1\,\mathrm{GeV})$ were considered together with a
light gluino.}
\cite{Pagels:1981ke,Goldberg:1983nd,Ellis:1983ew}. Such a photino
constitutes hot dark matter. If this would provide the entire dark
matter, it is inconsistent with structure formation in the early
universe
\cite{White:1984yj,Smoot:1992td,Spergel:2003cb,Abazajian:2005xn,Viel:2007mv}.
More recently it was found
\cite{Hooper:2002nq,Belanger:2002nr,Bottino:2002ry,Bottino:2003iu,Belanger:2003wb,Lee:2007ai,Profumo:2008yg}
that a light cold dark matter neutralino of mass ${\cal O}(5\,\mathrm
{GeV})$ can be obtained within the MSSM without Eq.~(\ref{assump}),
while still providing dark matter to be consistent with the WMAP
observations~\cite{Spergel:2006hy,Dunkley:2008ie}. As opposed to the
earlier work on a light photino, we consider here a light neutralino;
in fact we show that these are predominantly bino, given experimental
constraints.  Furthermore we consider significantly lighter, possibly
massless, neutralinos. A stable and light neutralino constitutes hot
dark matter. For $\mneu{1}\ll 25\,$eV it contributes only a small
amount to the energy density of the universe and should thus be
consistent with observations. We determine the precise number in
Sect.~\ref{cowsik-mcc}. Another dark matter candidate is then however
required \cite{Steffen:2008qp}. A light photino or neutralino of mass
${\cal O}(10\,\mathrm{MeV })$ could also be produced in supernova
explosions
\cite{Grifols:1988fw,Ellis:1988aa,Lau:1993vf,Dreiner:2003wh,Kachelriess:2000dz}.\footnote{For
  a particular application of light but massive photinos see also
  Refs.~\cite{Sciama:1982ij}.} We reinvestigate these astrophysical
and cosmological questions particularly for a massless neutralino in
Sect.~\ref{ch:cosmo}.

\medskip

As mentioned above, the main focus of this paper is to analyse in
detail the issue of laboratory constraints on a light neutralino.  We
consider collider searches for light neutralinos, the contribution of
light neutralinos to precision electroweak observables via radiative
corrections and neutralinos in rare meson decays.  At colliders, the
most promising processes for direct searches are
\begin{eqnarray}
e^++e^-&\rightarrow& \neu{1}\neu{1}\gamma\,,
\label{proc-1}\\
e^++e^-&\rightarrow& e^+e^-\neu{1}\neu{1}\,.
\label{proc-2}
\end{eqnarray}
These were investigated early for a photino
\cite{Fayet:1982ky,Ellis:1982zz}. Three of the present
authors~\cite{Dreiner:2006sb,Dreiner:2007vm} have recently presented
the first complete calculation of the process Eq.~(\ref{proc-1}) for a
general neutralino, focusing on heavy neutralinos, and analysed the
resulting phenomenology at the ILC including beam polarisation.  In
Sect.~\ref{ch:chi12}, we reanalyse this for the case of a massless
neutralino. We also consider the case of associated production of the
lightest and next-to-lightest neutralino~\cite{Choudhury:1999tn}.

\medskip

A second method to investigate a light neutralino in the laboratory is
via its virtual corrections to precision observables. A first point
that must be considered when discussing light neutralinos is whether
the light mass is stable under radiative corrections. The neutralino
also contributes to all electroweak observables via radiative
corrections. For early work see for example
Refs.~\cite{Chankowski:1993eu,deBoer:1996vq,Erler:1998ur}, and
Ref.~\cite{PomssmRep} for a recent review. In most analyses the
assumption Eq.~(\ref{assump}) was typically made; the case of varying
the parameter $M_1$ independently was not systematically analysed. In
general one would expect that a very light neutralino could give rise
to significant supersymmetric corrections to the precision
observables. We study this issue in Sect.~\ref{prec-observ}, where we
discuss in detail the total and invisible $Z$~widths, $\Gamma_Z$ and
$\Gamma_{\rm inv}$, the $W$~boson mass $\MW$, the effective leptonic
weak mixing angle $\sweff$, the anomalous magnetic moment of the muon
$(g-2)_\mu$ as well as electric dipole moments in the case of complex
parameters.

Concerning rare meson decays to neutralinos, the first process studied
in the literature
was~\cite{Ellis:1982ve,Gaillard:1982rw,Nieves:1985ir}
\begin{equation}
K\ra \pi \neu{1}\neu{1}\,.
\end{equation}
For a light neutralino the experimental signature is equivalent to the
SM decay $K\ra \pi \nu\bar\nu\,,$ which has since been observed
experimentally~\cite{Adler:1997am,Adler:2001xv,Anisimovsky:2004hr,Adler:2008zz}.
The branching ratio still has a large experimental error but is so far
consistent with the SM prediction~\cite{Buras:2005gr,Isidori:2005xm,Mescia:2007kn,Brod:2008ss}.  
This can thus be used to set bounds on the related supersymmetric
parameters. However, typically the decay involves virtual squarks and
if these are sufficiently heavy, no bound on the neutralino mass is
obtained. For a massless or near massless neutralino, we can consider
this instead as a lower bound on the relevant sfermion mass, as in the
original study of the anomalous magnetic moment of the
muon~\cite{Fayet:1974fj}. In Sect~\ref{meson-decays}, we present a
detailed overview of $K$-meson, pion, $B$-meson, and quarkonium decays
\cite{Gaillard:1982rw,Haber:1984rc,Dobroliubov:1987cb,Adhikari:1994wh,Adhikari:1994iv,Adhikari:1995bq,McElrath:2005bp,Chang:1997tq}.
A more complete treatment of the bounds resulting from meson decays,
including complete (higher order) calculations of the decay rates is
deferred to a separate paper~\cite{work-ip:2008}.

}

\section{The Neutralino Framework}
\label{framework}
\subsection{Tree-level and higher-order corrections}
\label{sec:tree}

We shall work in the minimal supersymmetric SM (MSSM). The part of
the Lagrangian which describes the neutralino mixing is given
by~\cite{Drees:2004jm}
 \begin{eqnarray}
\label{ul:eq:chimix}
\mathscr{L}_{\neu{}} &=& -\half \bino\bino  M_1 -\half \wino^0\wino^0  M_2 
		+ \mu \widetilde{h}_1^1\widetilde{h}_2^2 
		- \frac{g_2}{2}\wino^0(v_1 \widetilde{h}_1^1 - v_2 
\widetilde{h}_2^2) 
+ \frac{g_1}{2}\bino(v_1 \widetilde{h}_1^1 - v_2 \widetilde{h}_2^2)
\\[2mm]
	     &\equiv& - \half \psi^T_0 \mathcal{M}_{\neu{1}} \psi_0 ,
\end{eqnarray}
and the fermionic fields are two-component Weyl spinors
\cite{Dreiner:2008tw}. Here
\begin{eqnarray}
\label{neutralinomatrix}
\mathcal{M}_{\neu{}} &=&
\begin{pmatrix}
M_1 & 0   & - \MZ \sw \cos\beta & \phantom{-}\MZ\sw \sin\beta \\
0   & M_2 & \phantom{-} \MZ \cw \cos\beta  & -\MZ \cw\sin\beta \\
 - \MZ \sw \cos\beta &\phantom{-} \MZ \cw \cos\beta  & 0 & -\mu\\
\phantom{-}\MZ\sw \sin\beta& -\MZ \cw\sin\beta & -\mu & 0
\end{pmatrix}\,,
\end{eqnarray}
and
\begin{eqnarray}
\psi^T_0 &\equiv& 
\begin{pmatrix}\bino,& \wino^0,&\widetilde{h}_1^1,&\widetilde{h}_2^2
\end{pmatrix}\,.
\end{eqnarray}
$\MZ$ is the $Z$ gauge-boson mass and $\sw\equiv\sin\theta_
{\mathrm{w}},\,\cw\equiv\cos\theta_{\mathrm{w}}=\MW/\MZ$, where
$\theta_ {\mathrm{w}}$ is the weak mixing angle. $\mu$ is the higgsino
mass parameter and $\tan\beta \equiv v_2/v_1$ the ratio of the two
vacuum expectation values of the Higgs doublets.

The chargino mixing is described by the following matrix~\cite{Drees:2004jm}:
\begin{eqnarray}
\mathscr{L} &=& - (\psi^-)^T X \psi^+ , \qquad\qquad\;\;\; \mathrm{where} \quad
\psi^+ \equiv (\wino^+,\widetilde{h}_2^1)^T,\quad  \psi^- \equiv 
(\wino^-,\widetilde{h}^2_1)^T,
\\[2mm]
X & \equiv &
\begin{pmatrix}
M_2 & \sqrt{2} \MW \sin\beta\\[2mm]
 \sqrt{2} \MW \cos\beta & \mu
\end{pmatrix}\,.
\end{eqnarray}
The mass matrix $X$ is diagonalised by a biunitary transformation:
\begin{eqnarray}
\diag(\mcha{1} , \mcha{2}) = U^\ast X V^{-1}, 
\end{eqnarray}
with $U$, $V$ unitary $2\times 2$ matrices, see for example 
\cite{Haber:1984rc}. The lower experimental bound on the lightest 
chargino mass is~\cite{Amsler:2008zz}
\begin{equation}
\mcha{1} > 94\GeV\,.
\label{eq:bound1}
\end{equation}
Scanning over the parameter space, taking this bound into account,
leads to a lower bound on both $|\mu|$ and $M_2$ \cite{Barger:2005hb}
\begin{equation}
|\mu|,\, M_2 \gsim 100\GeV\,. 
\end{equation}
If the GUT relation between $M_1$ and $M_2$, Eq.~(\ref{assump}), is
assumed, then this implies 
\begin{equation}
M_1\gsim 50\,\GeV\,.
\label{eq:bound3}
\end{equation}
The usually quoted lower bounds on the lightest neutralino mass
arise from imposing the experimental bound on the lightest
chargino mass, Eq.~(\ref{eq:bound1}), in conjunction with the assumption
of the GUT relation between $M_1$ and $M_2$,
Eq.~(\ref{assump}). The value given in Eq.~(\ref{lep-bound}) furthermore
takes into account results from other searches for supersymmetric
particles and constraints from the Higgs sector, see
Ref.~\cite{Abdallah:2003xe} for details.

\begin{figure}[t]
\begin{picture}(200,180)
\put(15,-10){\includegraphics{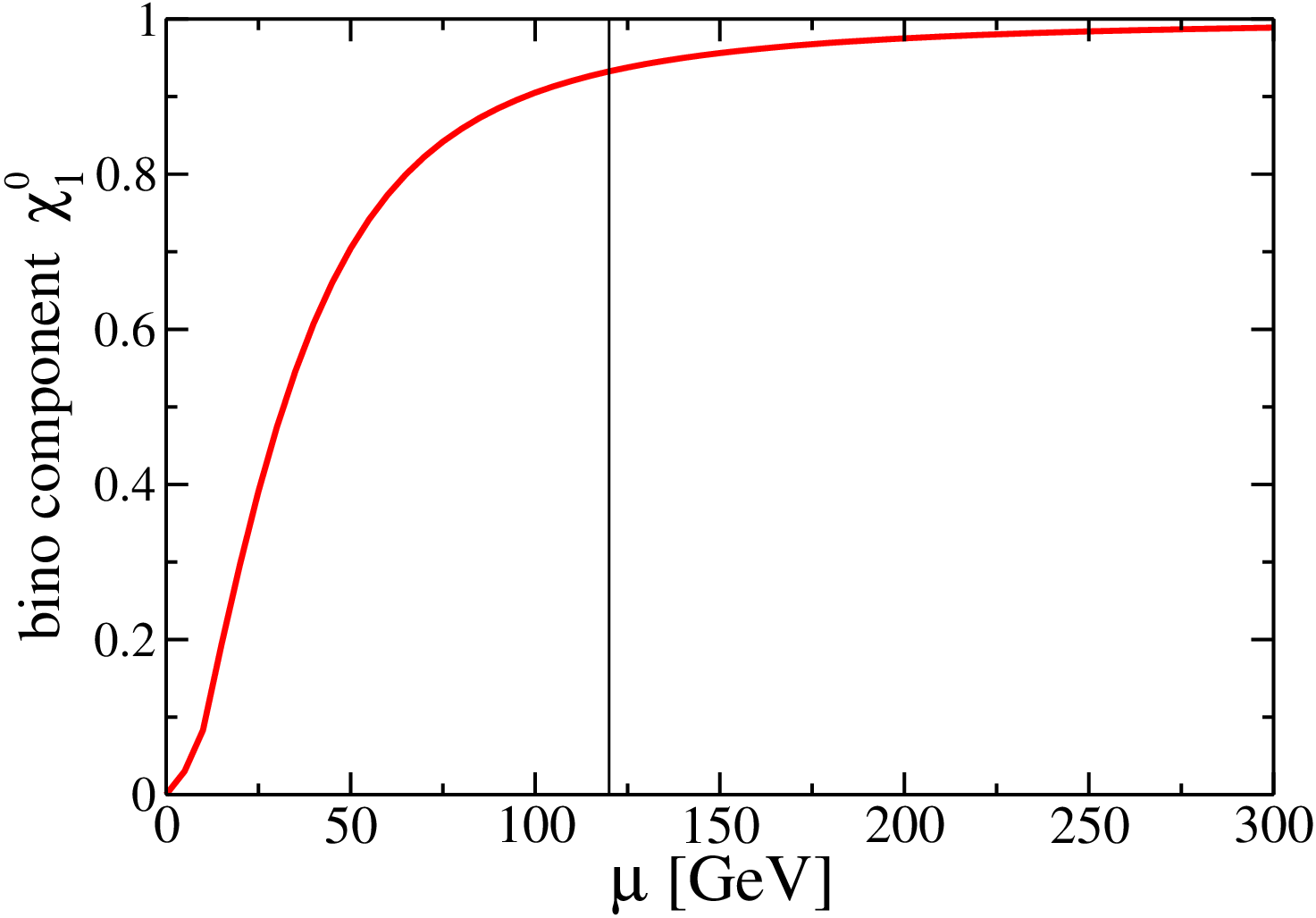}}
\put(250,-10){\includegraphics{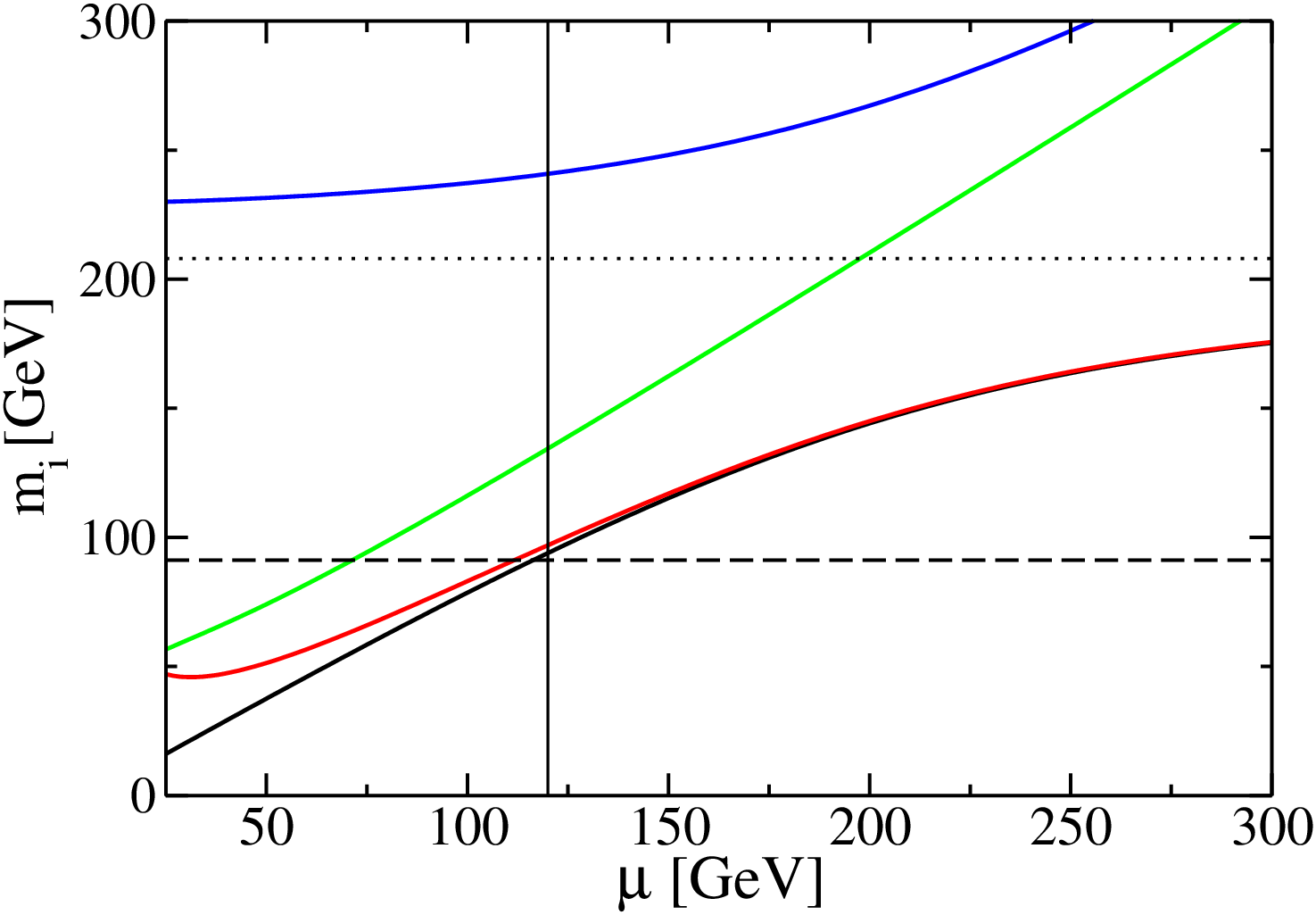}}
\put(49.50,30){\footnotesize $ m_{\tilde\chi_1^\pm} \!\!<\!94$~GeV}
\put(390,80){\footnotesize $\neu{2}$}
\put(415,110){\footnotesize $\neu{3}$}
\put(295,110){\footnotesize $\neu{4}$}
\put(295,25){\footnotesize $\cha{1}$}
\put(280,83){\footnotesize $\sqrt s= 208$~GeV}
\put(370,40){\footnotesize $\MZ\approx 91$~GeV}
\end{picture}
\caption{\small Bino admixture of $\neu{1}$ (left plot) and masses of
  charginos and neutralinos (right plot) for $M_2=200$~GeV,
  $\tan\beta=10$, and $M_1$ as given in
  Eq.~(\ref{massless-neut-condition}), such that
  $\mneu{1}=0$~\cite{Dreiner:2007fw}.  Left of the vertical lines at
  $\mu\approx120$~GeV, the chargino mass is $m_{\tilde\chi_1^\pm}<
  94$~GeV.  In the right panel, the dotted line indicates the
  kinematic reach of LEP2 ($\sqrt s= 208$~GeV) for $e^+e^- \to
  \neu{1}\neu{i}$ production ($i = 2, 3, 4$), and the dashed line
  indicates the mass of the Z boson, $\MZ\approx 91$~GeV. Note that
  $\tilde\chi^0_2$ is nearly mass degenerate with $\tilde\chi_1^\pm$
  for $\mu>120\,$GeV.  }
\label{fig:neutmixandmass}
\end{figure}

If instead the theoretical assumption of Eq.~(\ref{assump}) is dropped
and $M_1$ and $M_2$ are treated as independent free parameters, there is
an additional freedom in determining the lightest neutralino mass.
In fact we can require a vanishing
lightest neutralino mass at tree-level, by setting the determinant of
the mass matrix, Eq.~(\ref{neutralinomatrix}), to zero
\begin{equation}
\mathrm{det}(\mathcal{M}_{\tilde\chi^0})=0\,.
\end{equation}
This is equivalent to~\cite{Bartl:1989ms,Gogoladze:2002xp}
\begin{equation}
\mu\,\big[\,M_2 \MZ^2\sw[2]\sin(2\beta) + M_1\big(-M_2 \mu 
+ \MZ^2\cw[2]\sin(2\beta)\big)\,\big]=0\,.
\end{equation}
The solution $\mu=0$ is excluded by the above chargino bounds. Solving
for $M_1$ yields
\begin{eqnarray}
M_1 = \frac{M_2 \MZ^2 \sin(2\beta)\sw[2]}{\mu M_2 - \MZ^2\sin(2\beta)\cw[2]}\,.
\label{massless-neut-condition}
\end{eqnarray}
Accordingly, for fixed values of $\mu,\,M_2$ and $\tan\beta$
one can always find a value of $M_1$ such that the lightest
neutralino becomes massless. Typically, the second term in the
denominator is much smaller than the first, so that the resulting
expression for $M_1$ is approximately given by 
[making use of $\sin(2\beta)=2\tan\beta/(1+\tan ^2\beta)
\approx 2/\tan\beta$, for $\tan\beta \gsim 3 $]
\begin{eqnarray}
M_1 \approx \frac{\MZ^2 \sin(2\beta)\sw[2]}{\mu}\approx2.5\,\GeV 
\left(\frac{10}{\tan\beta}\right) \left(\frac{150\,\GeV}{\mu}\right)\,.
\label{massless-neut-value}
\end{eqnarray}
This typically implies $M_1\ll M_2, \mu$. In this parameter region
the lightest neutralino $\neu{1}$ is predominantly bino, \textit{
i.e.} it couples to hypercharge. The bino admixture is typically above
90\% in the parameter range where the chargino mass bound is
satisfied.  The masses of the other neutralinos and charginos are of
the order of $M_2$ and $\mu$, see Fig.~\ref{fig:neutmixandmass}.

\medskip

The results given in Eqs.~(\ref{massless-neut-condition}) and
(\ref{massless-neut-value}) have been obtained at the tree-level.
Since the chargino/neu\-tra\-lino sector is described by the three
independent parameters $M_1$, $M_2$ and $\mu$ (in the case of real
parameters, and we furthermore assume that $\tan\beta$ is determined
via the Higgs sector of the MSSM), only three of the six chargino and
neutralino masses are independent. Consequently, the other three
masses are predicted. When including radiative corrections, one has to
choose a certain renormalization scheme to define the physical meaning
of the parameters. The three independent parameters of the
chargino/neutra\-lino mass matrices can be traded for three masses
that are specified as input quantities. While in general loop
corrections can give rise to a shift between the physical mass and the
tree-level mass, the three masses chosen as independent input
parameters do not receive higher-order corrections by
construction. For the discussion of a very light neutralino it is thus
convenient to choose $\mneu{1}$ as one of the input parameters in
order to avoid that higher-order corrections could drive it away from
zero. Such a scheme where the two chargino masses and the lightest
neutralino mass have been renormalized on-shell has been applied for
the calculation of higher-order corrections in the MSSM
chargino/neutralino sector~\cite{Fritzsche:2002bi,Oller:2003ge}.
Accordingly, once the mass of the lightest neutralino has been
arranged to be small at tree-level, with an appropriate choice of
renormalization scheme it remains small also if higher-order
corrections are taken into account.  The same also holds in the case
of complex parameters which is discussed in the next subsection.


\subsection{Complex parameters}
\label{CPphases}

The condition for a massless neutralino can also be obtained for a
CP-violating neutralino sector.  Then the parameters
\begin{eqnarray}
\label{ul:eq:complex}
M_1 = |M_1|\eu^{\ie \phi_1}\qquad \text{and} \qquad 
\mu = |\mu|\eu^{\ie \phi_\mu} 
\end{eqnarray}
are complex and have CP-violating phases $\phi_1$ and $\phi_\mu$.  We
choose the convention where $M_2$ is real and positive, absorbing its
possible phase by a redefinition of the gaugino fields.  In the
presence of complex phases, two equations have to be separately
fulfilled to have a zero mass neutralino,
\begin{eqnarray}
\imag \{ \mathrm{det}(\mathcal{M}_{\neu{}}) \} = 0\,, 
\qquad {\rm and }\qquad 
\real \{ \mathrm{det}(\mathcal{M}_{\neu{}})\} = 0\,.
\label{eq:CPcase}
\end{eqnarray}
These conditions are equivalent to
\begin{eqnarray}
\MZ^2 \cw[2]\sin(2\beta) \sin\phi_1 
-|\mu| M_2 \sin(\phi_1 + \phi_\mu) &=&0,
\label{eq:firstCPcondition}\\[2mm]
 M_2 \MZ^2\sw[2]\sin(2\beta) + |M_1|\big[
-M_2 \mu \cos(\phi_1 + \phi_\mu)
+ \MZ^2\cw[2]\sin(2\beta) \cos\phi_1 \big]&=&0,
\label{eq:twoCPconditions}
\end{eqnarray}
respectively, and they can be solved for the absolute values of
\begin{eqnarray}
       |\mu| = \frac{\MZ^2 \cw[2] \sin(2\beta) \sin\phi_1}{M_2 \sin(\phi_1+\phi_\mu)}
 \qquad\text{and}\qquad
       |M_1| =  -M_2 \tw[2]\frac{\sin(\phi_1 + \phi_\mu)}{\sin\phi_\mu}\,.
\label{eq:CPcase1}
\end{eqnarray}
The equations can also be converted to any other set of two neutralino
parameters, for example
\begin{eqnarray}
M_2 & = & \frac{\MZ^2 \cw[2] \sin(2\beta) \sin\phi_1}{|\mu| \sin(\phi_1+\phi_\mu)}
      \qquad\text{and}\qquad
|M_1|  =   -\frac{\MZ^2 \sw[2] \sin(2\beta)\sin\phi_1}{|\mu| \sin\phi_\mu}.
\label{eq:CPcase2}
\end{eqnarray}
Note that in the presence of non-zero complex phases, the conditions 
for a massless neutralino cannot always be fulfilled.  For example, 
it follows from
Eq.~(\ref{eq:CPcase2}) that the phases have to fulfill $\sin\phi_1/
\sin(\phi_1+\phi_\mu) > 0$ and $\sin\phi_1/\sin\phi_\mu<0$. 
In the CP-conserving limit with vanishing phases,
we retrieve the condition Eq.~(\ref{massless-neut-condition}) 
for $M_1$ for a massless neutralino 
from Eq.~(\ref{eq:twoCPconditions}), with
$ \cos(\phi_1 + \phi_\mu) \to 1$,
$\cos\phi_1 \to 1$, and Eq.~(\ref{eq:firstCPcondition}) is trivially fulfilled.

\section{Collider Bounds}
{
\input paperdef

In this section, we consider the bounds on a light neutralino from
collider searches. We focus our discussion in particular on the direct
searches performed by the experimental collaborations at LEP. The
bounds from LEP are in general more stringent than the ones from other
lepton colliders with lower energies and/or luminosities. We briefly
discuss limits from $b$-factories in \refse{radiative}.  Concerning
bounds from hadron colliders, at the Tevatron the large QCD background
limits the sensitivity in the search for direct production of a light
or nearly massless neutralino~\cite{Adams:2008nu}. At the LHC there
could be better prospects for detecting effects of a very light
neutralino in cascade decays of other (heavy) SUSY
particles~\cite{Bottino:2008xc}.

\subsection{Neutralino production at LEP }
\label{ch:chi12}

If we assume a massless neutralino $\neu{1}$ by choosing $M_1$ as
given in Eq.~(\ref{massless-neut-condition}), the mass of the
next-to-lightest neutralino $\neu{2}$ is mainly determined by the
values of $\mu$ and $M_2$.  In Fig.~\ref{fig:nm2sp12}(a), we show 
contour lines of the second lightest neutralino
mass.\footnote{
Note that here and for the following scenarios we choose 
an intermediate value of  $\tan\beta=10$ and discuss $\mu>0$ only, since
the masses and cross sections 
change only slightly of the order of $10\%$ if we take larger values
of $\tan\beta$, and/or negative values of $\mu$.}
Qualitatively, the dependence is $\mneu{2}\sim M_2 $
for $M_2 \ll \mu$, and similarly $\mneu{2}\sim \mu $ for $M_2 \gg
\mu$.  This can be also observed in Fig.~\ref{fig:neutmixandmass},
where we show the dependence of the masses on $\mu$.  Thus for $\mu$
or $M_2 \lsim 200$~GeV, the associated production of neutralinos,
$e^+e^-\to\neu{1}\neu{2}$, \textit{cf.} Fig.~\ref{fig:FeynNeutprod},
would be accessible at LEP up to the kinematical limit of $\sqrt
s=\mneu{2} = 208 \GeV$, if $\mneu{1} =0$. 

In order to compare with the results of the LEP searches we make use
of the model-independent upper bounds on the topological neutralino
production cross section obtained by the OPAL collaboration in the
searches at LEP with $\sqrt s= 208$~GeV~\cite{Abbiendi:2003sc},
\begin{equation}
\sigma(e^+e^-\to\neu{1}\neu{2})\times
{\rm BR}(\neu{2}\to Z\neu{1})\times {\rm BR}(Z\to q\bar q) .
\end{equation}
In this analysis 
a stable lightest neutralino is assumed. In Fig.~\ref{fig:nm2sp12}(b)
the observed limit at $95\%$ confidence level on 
$\sigma(e^+e^-\to\neu{1}\neu{2})\times {\rm BR}(\neu{2}\to Z\neu{1})$ is
shown (taking into account that ${\rm BR}(Z\to q\bar q)\approx 70\%$)
in the $\mneu{1}$--$\mneu{2}$ plane~\cite{Abbiendi:2003sc}. For
$\mneu{1}=0$ (and $\mneu{2} \lsim 190 \gev$) one can roughly read off
the upper limit 
\begin{equation}
\sigma(e^+e^-\to\neu{1}\neu{2}) \times {\rm BR}(\neu{2}\to Z\neu{1})
<70\,\mathrm{fb}\,.  
\end{equation}

\begin{figure}[t]
\begin{picture}(200,180)
\put(0,-30){\includegraphics{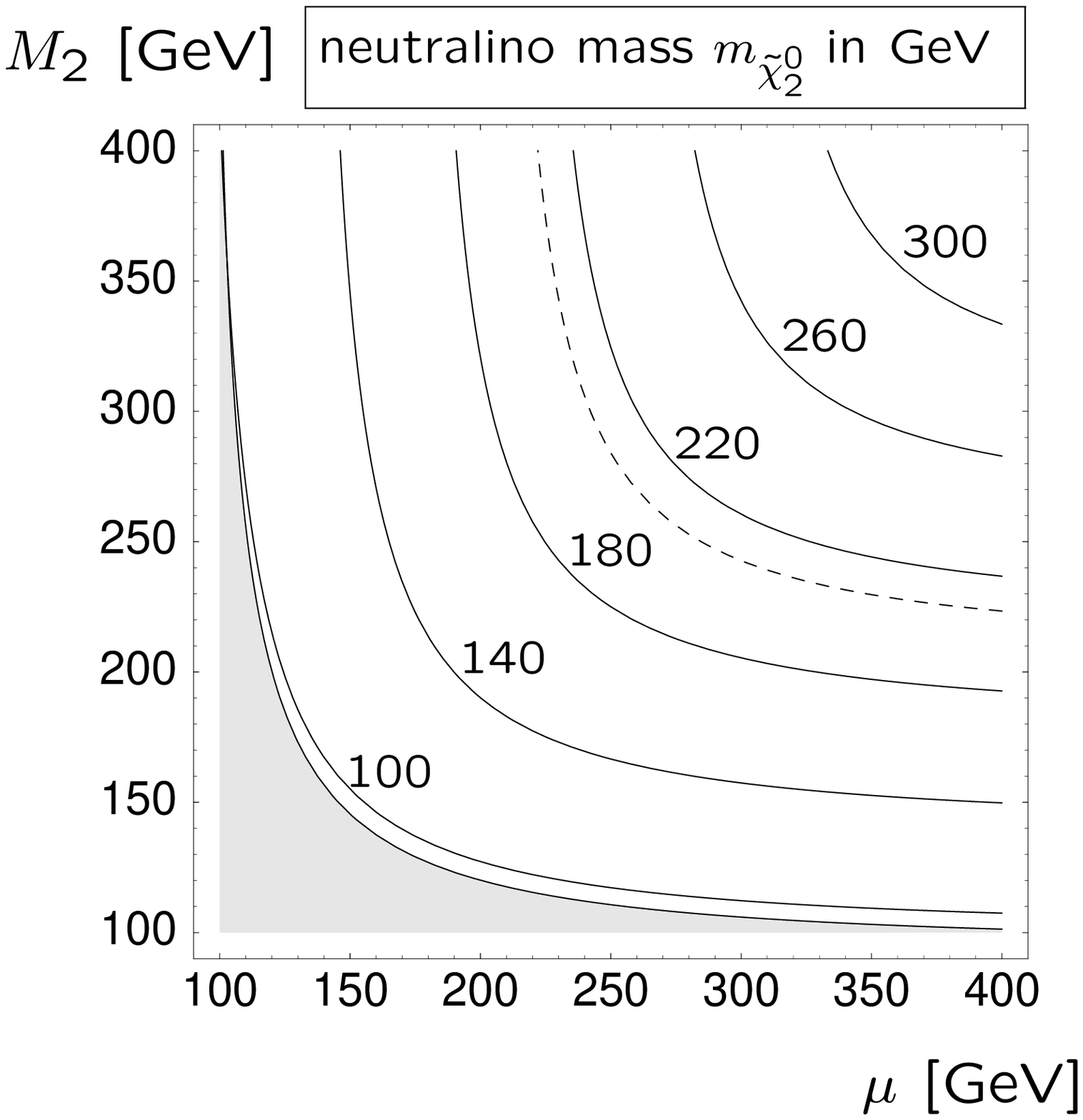}}
\put(230,-30){\includegraphics{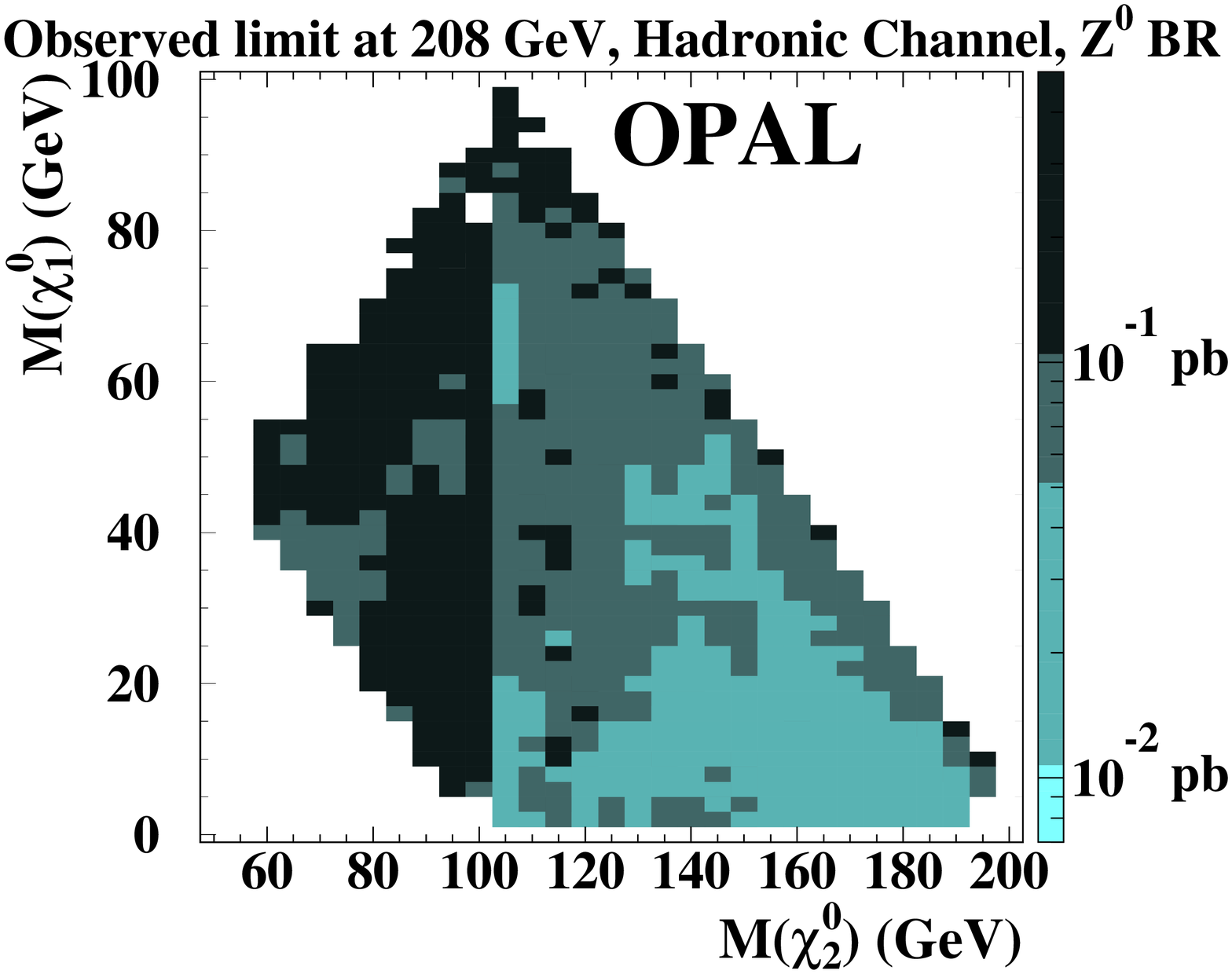}}
\put(50,0){\small\bf (a)}
\put(280,0){\small\bf (b)}
\end{picture}
\caption{\small {\bf (a)} Contour lines in the $\mu$--$M_2$ plane of
  the neutralino mass $\mneu{2}$. In the grey shaded area the chargino
  mass is $m_{\tilde\chi_1^\pm}< 94$~GeV.  The dashed line indicates
  the kinematical limit $\mneu{2}= \sqrt s=208$~GeV at LEP2.
  Throughout we have chosen $M_1$ such that $\mneu{1}=0$. The
  lightest chargino is nearly mass degenerate with $\tilde\chi^0_2$
  for $M_2\gsim200\GeV$ and $\mu\gsim125\GeV$.  {\bf (b)} $95\%$
  confidence limit on the cross section
  $\sigma(e^+e^-\to\neu{1}\neu{2})\times {\rm BR}(\neu{2}\to
  Z\neu{1})$ at $\sqrt s = 208$~GeV (taken from
  Ref.~\cite{Abbiendi:2003sc}, Fig.~10).  }
\label{fig:nm2sp12}
\end{figure}

\begin{figure}[t]
\hspace{-2.cm}
\begin{minipage}[t]{6cm}
\begin{center}
{\setlength{\unitlength}{0.6cm}
\begin{picture}(5,5)
\put(-2.5,-8.5){\includegraphics{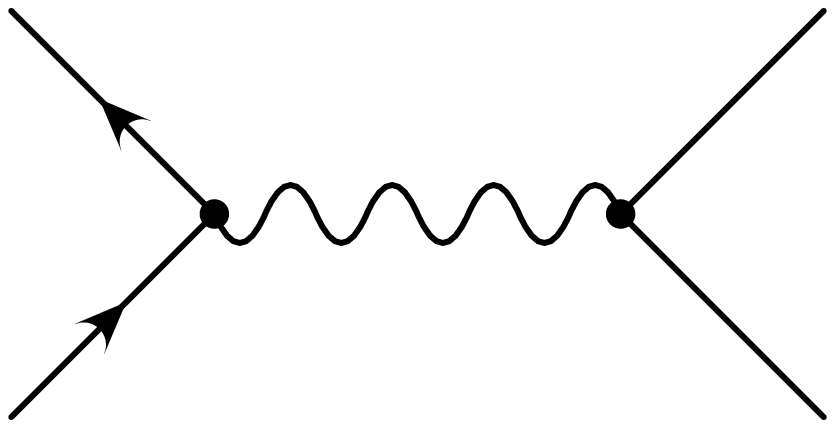}}
\put(1.7,-.4){{\small $e^{-}$}}
\put(7.8,-.4){{\small $\tilde\chi^0_j$}}
\put(1.7,3.8){{\small $e^{+}$}}
\put(7.8,3.8){{\small $\tilde\chi^0_i$}}
\put(5.1,2.5){{\small $Z^0$}}
\end{picture}}
\end{center}
\end{minipage}
\hspace{-0.5cm}
\begin{minipage}[t]{5cm}
\begin{center}
{\setlength{\unitlength}{0.6cm}
\begin{picture}(2.5,5)
\put(-4,-9){\includegraphics{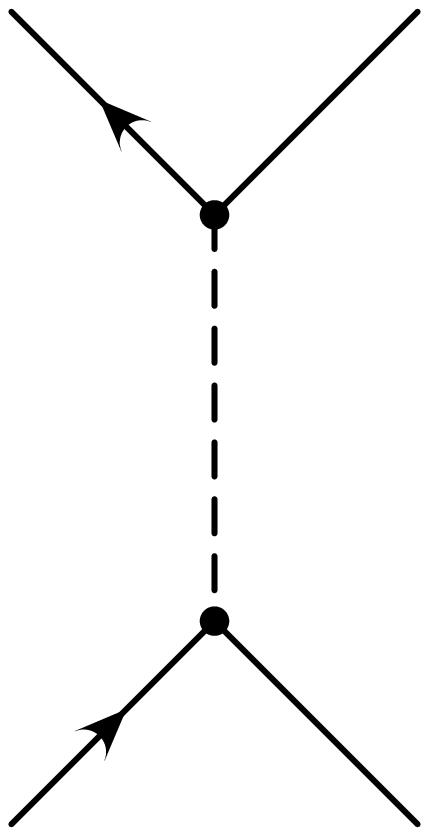}}
\put(1.8,-1.5){{\small $e^{-}$}}
\put(1.8,3.8){{\small $e^{+}$}}
\put(5.8,-1.5){{\small $\tilde\chi^0_j$}}
\put(5.8,3.8){{\small $\tilde\chi^0_i$}}
\put(4.4,1.5){{\small $\tilde{e}_{L,R}$}}
 \end{picture}}
\end{center}
\end{minipage}
\begin{minipage}[t]{5cm}
\begin{center}
{\setlength{\unitlength}{0.6cm}
\begin{picture}(2.5,5)
\put(-4.5,-9){\includegraphics{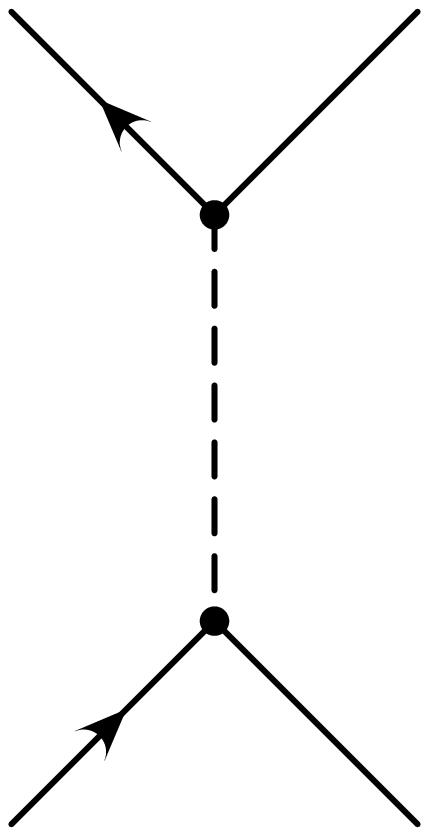}}
\put(1.3,-1.5){{\small $e^{-}$}}
\put(1.3,3.8){{\small $e^{+}$}}
\put(5.5,-1.5){{\small $\tilde\chi^0_i$}}
\put(5.5,3.8){{\small $\tilde\chi^0_j$}}
\put(3.9,1.5){{\small $\tilde{e}_{L,R}$}}
\end{picture}}
\end{center}
\end{minipage}
\vspace{.7cm}
\caption{\label{fig:FeynNeutprod}Feynman diagrams for neutralino production
  $e^{+}e^{-}\to\tilde{\chi}^0_i\tilde{\chi}^0_j$
.}
\end{figure}

We analyze this bound assuming conservatively that ${\rm
BR}(\neu{2}\to Z\neu{1}) = 1$. In general this branching ratio can be
significantly smaller than 100\% since other decay modes like
$\neu{2}\to h \neu{1}$ and $\neu{2}\to e^\mp\tilde e^\pm$ (see below)
can be open. Imposing the bound $\sigma(e^+e^-\to\neu{1}\neu{2})
<70\,\mathrm{fb}$ significantly constrains the parameter space, since
the typical neutralino production cross sections are of the order of
$100$~fb for light neutralino and selectron masses. In
Fig.~\ref{fig:OPALbounds}(a) we show contour lines of the cross
section $\sigma(e^+e^- \to \neu{1}\neu{2})$ in the $\mu$--$M_2$ plane
for $\tan\beta = 10$ and degenerate selectron masses $m_{\tilde
e_R}=m_{\tilde e_L}=m_{\tilde e}=200$~GeV.  We observe that there is a
large region in the $\mu$--$M_2$ plane where the cross section is
higher than $70$~fb. In order to fulfill the bound, the selectron
masses have to be sufficiently heavy. It should be noted that the
bino-like $\neu{1}$ couples preferably to the $\tilde{e}_{R}$, which
is exchanged in the $t$ and $u$ channels, see the Feynman diagrams in
Fig.~\ref{fig:FeynNeutprod}.  Thus, the bound on the neutralino
production cross section can be translated into lower bounds on the
selectron mass $m_{\tilde e_R}=m_{\tilde e_L}=m_{\tilde e}$, for
$\mneu{1}=0$.  In Fig.~\ref{fig:OPALbounds}(b), we show contours of
the selectron mass, such that the bound
$\sigma(e^+e^-\to\neu{1}\neu{2}) < 70$~fb is fulfilled.  For example,
for a fixed selectron mass of $m_{\tilde e}=200$~GeV, the area below
the $200$~GeV contour in Fig.~\ref{fig:OPALbounds}(b) has a cross
section in excess of 70~fb.

\begin{figure}[t]
\begin{picture}(200,180)
\put(20,-0){\includegraphics{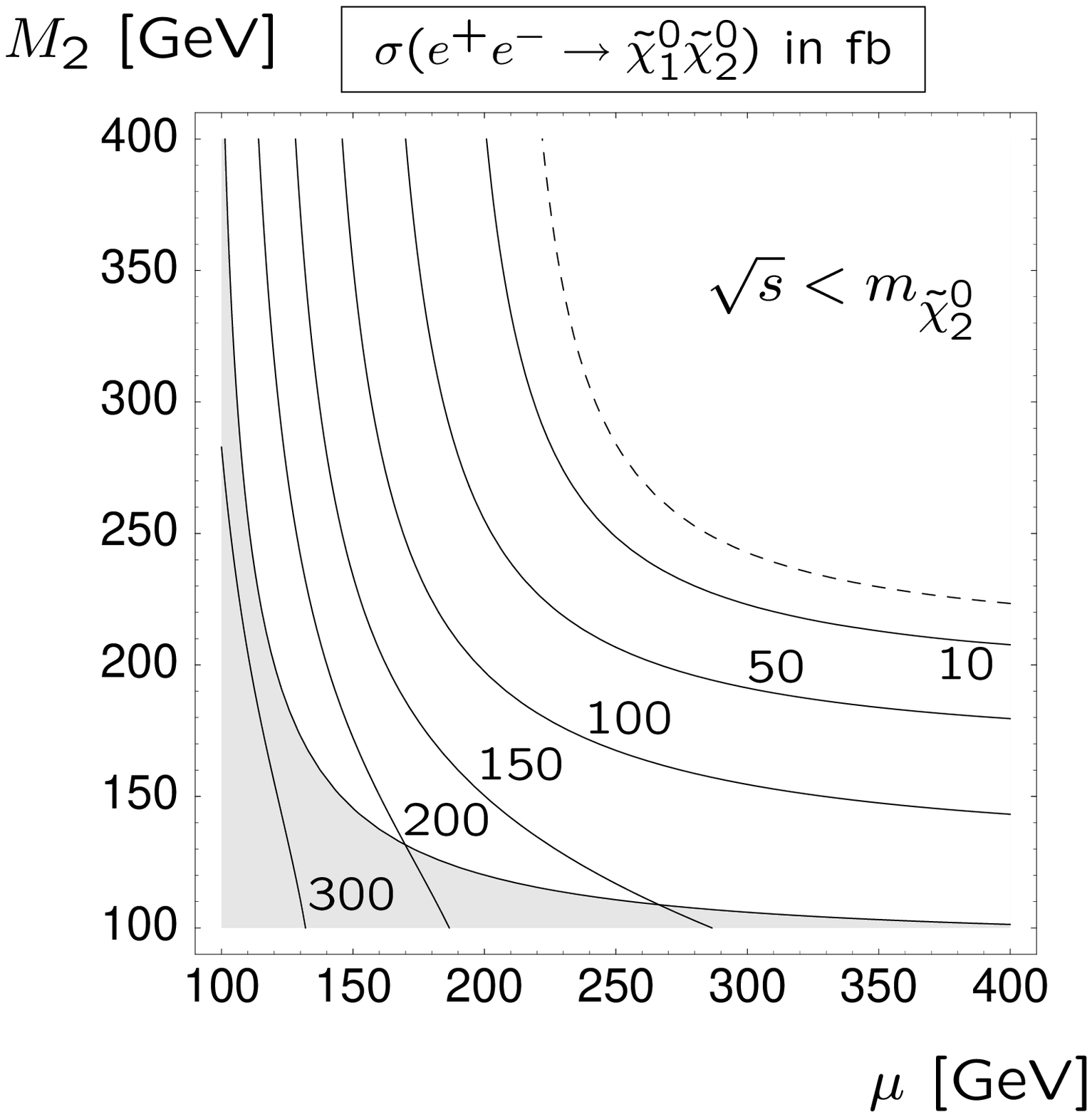}}
\put(230,-30){\includegraphics{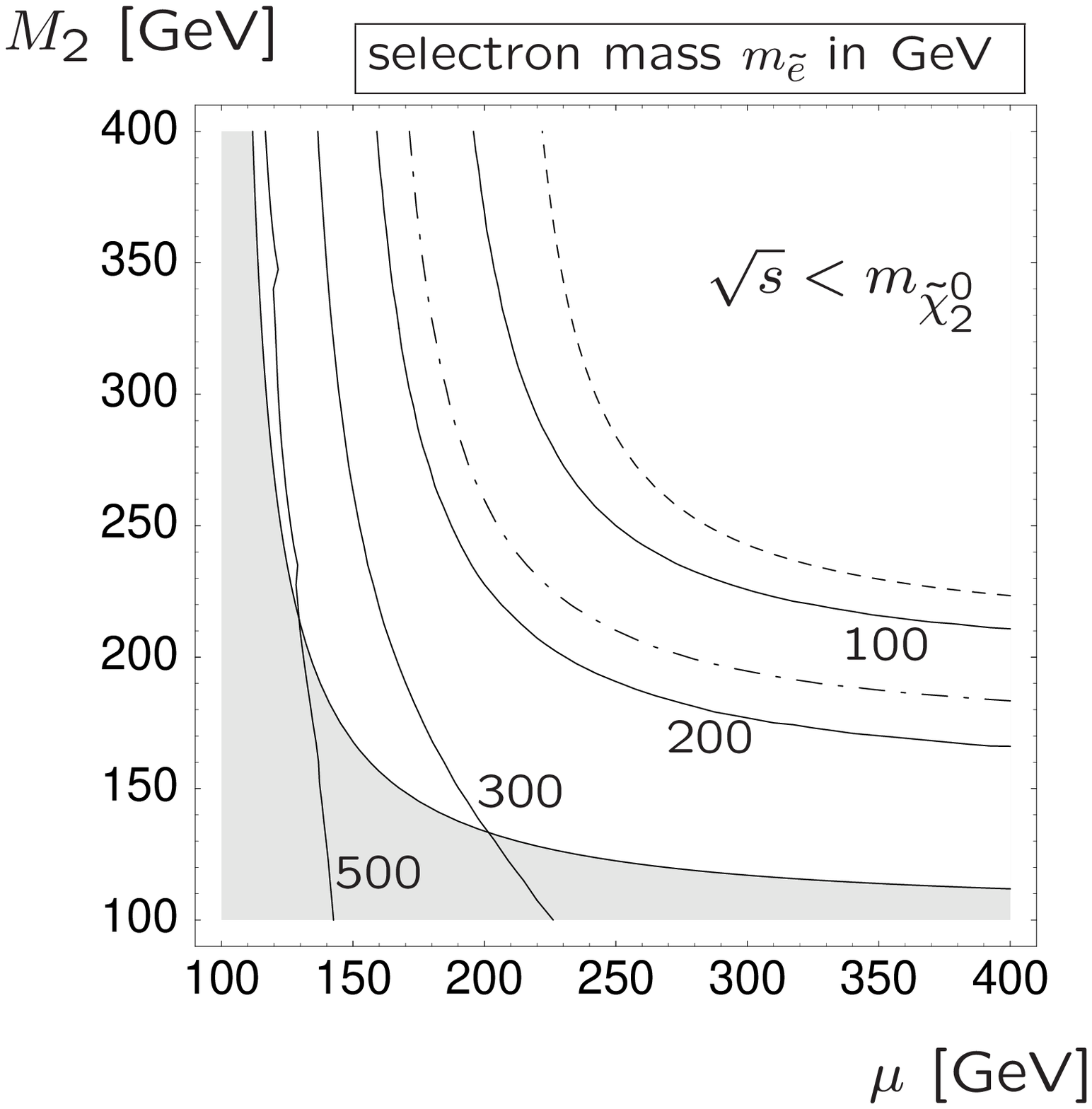}}
\put(50,0){\small\bf (a)}
\put(280,0){\small\bf (b)}
\end{picture}
\caption{\small {\bf (a)} Contour lines in the $\mu$--$M_2$ plane of
  the neutralino production cross section $\sigma(e^+e^-\to\neu{1}
  \neu{2})$ with $\tan\beta=10$, and $m_{\tilde e_R}=m_{\tilde e_L}=
  m_{\tilde e}=200$~GeV, at $\sqrt s = 208$~GeV.  At each point, $M_1$
  is chosen such that $\mneu{1}=0$.  {\bf (b)} Contour lines in the
  $\mu$--$M_2$ plane of the lower bounds on the selectron mass
  $m_{\tilde e_R}=m_{\tilde e_L}=m_ {\tilde e}$, such that
  $\sigma(e^+e^-\to\neu{1}\neu{2})=70$~fb for $\mneu{1}=0$
  with $\tan\beta=10$.  In {\bf (a}), {\bf (b)}, the dashed lines
  indicate the kinematical limit $\mneu{2}= \sqrt s=208$~GeV, in the
  grey shaded areas the chargino mass is $m_{\tilde\chi_1^\pm}<
  94$~GeV.  Along the dot-dashed contour in {\bf (b)} the relation
  $m_{\tilde e}=\mneu{2}$ holds. }
\label{fig:OPALbounds}
\end{figure}

It should be noted that above the dot-dashed contour in
Fig.~\ref{fig:OPALbounds}(b), the selectron is lighter than the
neutralino $\neu{2}$. Thus, in that region the two-body decay into a
selectron and electron, $\neu{2}\to e^\mp\tilde e^\pm$, is open,
leading to a reduction of ${\rm BR}(\neu{2}\to Z\neu{1})$.  The decay
of the second-lightest neutralino into a selectron and electron is
followed by $\tilde e^\pm\to e^\pm \neu{1}$.  The reduction of ${\rm
BR}(\neu{2}\to Z\neu{1})$ implies a decrease of the lower bound on the
selectron mass in this region. This effect is of minor relevance,
however, since the bound on the selectron mass in this region is
already close to the absolute lower experimental bound on the
selectron mass, $m_{\tilde e}\approx 75$~GeV~\cite{Amsler:2008zz}.  It
is clear from our analysis that for a sufficiently heavy selectron
mass a massless neutralino could not have been discovered at LEP.
Already for $m_{\tilde e}=200$~GeV, we have a significant range of
allowed parameter space, as one can see in
Fig.~\ref{fig:OPALbounds}(b).

\medskip

Finally we want to note that the bound on the neutralino production
cross section that we have used is indeed rather conservative. Since
the decay $\neu{3}\to Z\chi_1 ^0$ would also lead to the final state
$e^+ e^-\to q\overline{q}+\ssl{ E}$, the associated production of the
next heavier neutralino, $e^+e^-\to\neu{1}\neu{3}$, would increase the
signal rate if $\mneu{3}< \sqrt s$.  Since we have approximately
$\mneu{3}\approx\mu$, see \textit{e.g.} Fig~\ref{fig:neutmixandmass},
the reaction $e^+e^-\to\neu{1}\neu{3}$ would have been kinematically
accessible for $\mu\lsim 200$~GeV at LEP2 with a center-of-mass energy
of $\sqrt s=208$~GeV. Including the additional cross section from
$e^+e^-\to\neu{1}\neu{3}$ production in our analysis, we find that for
$\mu \lsim 150$~GeV the selectrons must now be heavier than $1$~TeV.
For $\mu\gsim200$~GeV there is no additional bound.

\subsection{Radiative neutralino production}
\label{radiative}

An additional search channel at LEP for a light neutralino would be
radiative neutralino production, $e^+e^-\to\neu{1}\neu{1}\gamma$.
However due to the large background from radiative neutrino
production, $e^+e^-\to\nu\bar\nu\gamma$, we find that the
(theoretical) significance\footnote{We define the significance $S$ as
the number of signal events over the square root of the number of
background events.} is at best $S\approx0.1$ for $\mathcal
L=100$~pb$^{-1}$ and $\sqrt
s=208\,$GeV~\cite{Dreiner:2006sb,Dreiner:2007vm}. In addition, cuts on
the photon energy or angle do not enhance the significance, due to
similar kinematic distributions of signal and background.  We find a
similar situation at $b$-factories, which are operating at the
$b$-meson resonances, $\sqrt s \approx 10$~GeV.  For example, with
$\mathcal L=100$~fb$^{-1}$ and $m_{\tilde e}=200$~GeV, we find a
significance of $S<0.1$, and signal-to-background ratios not larger
than $1\%$.  Further, an identification of the signal `photon plus
missing energy' is difficult due to the large photonic background from
the abundant hadronic processes at the $b$-factories~\cite{steve}.
Note that at the ILC, however, radiative neutralino production would
be measurable, due to the option of polarised beams, which allows to
reduce the background and enhance the signal at the same
time~\cite{Choudhury:1999tn,Dreiner:2006sb,Dreiner:2007vm,MoortgatPick:2005cw}.

}

\section{Precision observables}
\label{prec-observ}
{
{
\input paperdef
\renewcommand{\msusy}{M_{\rm SUSY}}
\renewcommand{\phiMe}{\cphi_1}

Electroweak observables have in the last decades played a key role in
constraining the free parameters of the SM and the MSSM (see
\textit{e.g.} Ref.~\cite{PomssmRep} for a recent review in the context
of the MSSM). In the following we study the impact of a light or
massless neutralino on electroweak precision physics. Among the key
observables in the electroweak sector are the mass of the $W$~boson~$
\MW$, the effective leptonic weak mixing angle~$\sweff$ (both 
discussed in~\refsubse{subse:MWsweff}), the electric dipole moments of
the electron, the neutron, and the mercury atom, and the anomalous
magnetic moment of the muon~$(g-2)_\mu$ (for the latter
see~\refsubse{subse:EDMgm2}). The total $Z$~boson decay width
$\Gamma_Z$ and the invisible $Z$~decay width $\Gamma_{\textup{inv}}$
are potentially very sensitive to a massless neutralino, as in
such scenarios the additional decay channel $Z\to\neu{1}\neu{1}$ is
kinematically allowed.  The resulting tree-level constraints on a
light neutralino were first investigated in
Refs.~\cite{Choudhury:1999tn,Dreiner:2003wh}. In a first step, in
\refse{subse:GammaZ} we reanalyse the impact of a massless neutralino
on the total width and the invisible width of the $Z$ boson, including
full one-loop and leading higher-order corrections.


\subsection{Total  \boldmath{$Z$}~Width  \boldmath{$\Ga_Z$} and Invisible
  \boldmath{$Z$}~Width~\boldmath{$\Ga_{\textup{inv}}$}}
\label{subse:GammaZ}
A light neutralino with mass $\mneu{1} \lsim \MZ/2$ can give 
contributions to the total and the invisible width of the $Z$~boson, 
in addition to the decay channels into SM fermions
\begin{equation}
\Ga_Z = \underbrace{\Ga_{Z, {\rm SM}}}_{\textup{decay into  SM fermions}} 
                       +\; \Ga_{\neu{1}}\,,  
\end{equation}
\begin{equation}
\Ga_{\textup{inv}} = 
  \underbrace{\Ga_{\textup{inv,SM}}}_{\textup{decay into neutrinos}} 
                       + \;\Ga_{\neu{1}}.
\end{equation}
Potentially the additional contributions due to $\Ga_{\neu{1}}$ can be
large if the neutralino has a considerable non-bino like component,
\textit{i.e.} a sizable coupling to the $Z$~boson. In
\citere{ZObsMSSM}, the processes~$Z\to\neu{1}\neu{1}$ and 
$Z\to f \bar f$ have been calculated at~\order{\al} and supplemented
with leading higher-order terms from the SM and the MSSM (see also the
discussion below). The corresponding results, which are the currently
most accurate MSSM predictions for these quantities, are used in the
following to analyse the impact of a massless neutralino on the
$Z$~decay width. The experimental values for the total width and the
invisible width of the $Z$ boson are~\cite{LEPEWWG,Amsler:2008zz}
\BEA
\label{GaZexp}
\Ga_Z^{\rm exp} &=& 2495.2 \pm 2.3 \mev~, \\
\label{GaZinv}
\Ga_{\rm inv}^{\rm exp} &=& 499.0 \pm 1.5 \mev~.  
\EEA 
Below, we label the experimental errors of these two quantities as
$\si_{\Ga_Z}^ {\rm exp}$ and $\si_ {\Ga_{\rm inv}}^{\rm exp}$,
respectively. In our numerical analysis, we show the results for
\begin{align}
\label{deGaZinv}
\de\Ga_{\rm inv} &\equiv \Ga_{\rm inv} - \Ga_{\rm inv}^{\rm exp}\,, \\
\label{deGaZ}
\de\Ga_Z &\equiv \Ga_Z - \Ga_Z^{\rm exp} ,
\end{align}
\textit{i.e.} the difference of the MSSM
prediction and the experimental result
for the invisible $Z$~width, $\de\Ga_{\textup{inv}}$, and the total
$Z$~width, $\de\Ga_Z$.

\medskip

In the following, we investigate both $\de\Ga_{\rm inv}$ and
$\de\Ga_Z$ in two representative SUSY parameter regions. As a first
scenario, we choose fairly light scalar fermions and set the diagonal
soft SUSY-breaking parameter $\msusy$ to $250\gev$ (in this section we
choose this parameter to be equal for all sfermions).  In
\reffi{fig:MUEM2plane}, we show $\de\Ga_{\rm inv}$ in the upper and
$\de\Ga_Z$ in the lower plot, as a function of $M_2$ and $\mu$. $M_1$
is fixed\footnote{A negligibly small offset value is added to $M_1$ to
  acquire numerical stability, while scanning the $\mu$-$M_2$-plane.}
via Eq.~(\ref{massless-neut-condition}).  The remaining SUSY
parameters are $\tb=10$, $A_{\tau}=A_t=A_b=\mgl=\MA= 500\gev$.  Here
$A_f$ ($f = t, b, \tau$) denotes the trilinear couplings of the
Higgses to the sfermions, $\mgl$ is the gluino mass, and $\MA$ denotes
the mass of the CP-odd Higgs boson. The deviations from the
experimental central values as given in Eqs.~(\ref{deGaZinv}),
(\ref{deGaZ}), are indicated as experimental $n\times\sigma$~contours
of the respective observable.
\begin{figure}[htb!]
\begin{center}
\inclMathplotB{DeltaGammaZInv.eps}\\
\vspace{1em}
\inclMathplotB{DeltaGammaZ.eps}
\end{center}
\caption{The difference of the experimental value and the theory 
  prediction for the invisible $Z$~width, $\de\Ga_{\rm inv}$, (upper
  plot) and the total $Z$~width, $\de\Ga_Z$, (lower plot) in the
  $\mu$--$M_2$-plane, both including the process $Z\to\neu{1}\neu{1}$.
  Deviations of the theory predictions from the experimental data are
  indicated as $\de\Ga_{\textup{inv}}\equiv(\Ga_{\textup{inv}}-\Ga^{
  \textup{exp}}_{\textup{inv}}) = (10,5,3,2,1)\times\sigma^{\textup{
  exp}}_{\Ga_{\textup{inv}}}$ (upper plot) and $\de\Ga_{Z}\equiv(\Ga
  _{Z}-\Ga^{\textup{exp}}_{Z}) = (20,10,3,2,1,0)\times\sigma^{\textup
  {exp}}_{\Ga_Z}$ (lower plot) contours. The SUSY parameters were 
  fixed as $\tb=10$, $\msusy = 250\gev, A_{\tau}=A_t=A_b=\mgl=\MA=500
  \gev$. For $M_1$ we use  \refeq{massless-neut-condition} (see text).}
\label{fig:MUEM2plane} 
\vspace{-2em}
\end{figure}
In addition, the $95\%\,$C.L.\ exclusion bounds of $\mcha{1}>94\gev$
\cite{Amsler:2008zz} on the chargino mass from direct searches are
marked by dashed white lines. The allowed regions are towards larger
values of $|M_2|$ and $|\mu|$. To the left, the plots stop at around
$\mu\approx -230\gev$, where the square of the lighter scalar top mass
turns negative. Fig.~\ref{fig:MUEM2plane} clearly displays that for
both observables the MSSM prediction can deviate considerably from the
experimental values. This is in particular the case for small $|\mu|$
and small $|M_2|$. Nearly all of the parameter space ruled out at the
$5\sigma$~level for $\Ga_{\textup{inv}}$ and at the $3\sigma$~level
for $\Ga_Z$ is, however, already excluded due to direct chargino
searches. For the interpretation of these plots it is furthermore
important to keep in mind that the results for $\Ga_{\textup{inv}}$
and $\Ga_Z$ do not only depend on $\mu$ and $M_2$, but on all the
other SUSY parameters as well. This means in particular that an
apparent $1 \sigma$ effect can easily be caused or canceled out by,
for instance, a change induced by $\msusy$, the mass scale of the
sfermions, which is known to have a strong impact on the decay into SM
fermions~(see also the discussion in Ref.~\cite{ZObsMSSM}).
Furthermore even in the SM, $\Ga_{\textup{inv}}$ is predicted to be
slightly larger than the experimentally measured value, resulting in a
$\sim 1 \si$ deviation.

\begin{figure}[htb!]
\begin{center}
\inclMathplotB{DeltaGammaZInv2.eps}\\
\vspace{1em}
\inclMathplotB{DeltaGammaZ2.eps}
\end{center}
\vspace{-2em}
\caption{Difference of experimental value and theory prediction for invisible
  $Z$~width, $\de\Ga_{\rm inv}$, (upper plot) and total $Z$~width,
  $\de\Ga_Z$, (lower plot) in the $\mu$--$M_2$-plane, both including
  the process $Z\to\neu{1}\neu{1}$.  Deviations of the theory
  predictions from the experimental data are indicated as
  $\de\Ga_{\textup{inv}}\equiv( \Ga_{\textup{inv}}-
\Ga^{\textup{exp}}_{\textup{inv}}) = 
(10,5,3,2,1.5,1)\times\sigma^{\textup{exp}}_{\Ga_{\textup{inv}}}$
(upper plot) and $\de\Ga_{\textup{inv}}\equiv( \Ga_Z-
\Ga^{\textup{exp}}_{Z}) = 
(20,10,3,2,1,0,-1)\times\sigma^{\textup{exp}}_{\Ga_Z}$
(lower plot) contours. 
The SUSY parameters are chosen as $\tb=10$, 
$\msusy = A_{\tau}=A_t=A_b=\mgl=\MA=600\gev$. $M_1$ is defined via
\refeq{massless-neut-condition} (see text).}
\label{fig:MUEM2plane2} 
\end{figure}

\medskip

As a second parameter range, we consider heavier sfermions: $\msusy
=600\gev$. The remaining parameters are given by $\tb=10$, $A_{\tau}=
A_t=A_b=\mgl=\MA=600\gev$. As before, we show in
\reffi{fig:MUEM2plane2} the results for the invisible $Z$ width
$\Ga_{\textup{inv}}$~(upper plot) and the total $Z$~width~$\Ga_Z$
(lower plot), as a function of $M_2$. $M_1$ is fixed in the same way
as before. Here the plot extends beyond $\mu=- 600\gev$.  The theory
predictions are again confronted with their experimental values as
described above. One finds somewhat better agreement of the $\Ga_Z$
predictions with the experimental data. This is mainly due to the
higher mass scale of the scalar fermions, which leads to the
prediction of lower $\Ga_Z$ values for a considerable part of the SUSY
parameter space (see Ref.~\cite{ZObsMSSM}). Despite the presence of
$\Ga_{\neu{1}}$, the overall prediction for $\Ga_Z$ even reaches
values below the experimental $1\sigma$ range, as indicated by the
$0\sigma$ and $-1\sigma$ contours. $\Ga_{\textup{inv}}$ is in slightly
better agreement with the data than before, but a $1\sigma$~deviation
can be observed for the entire allowed $\mu$-$M_2$-plane.

\medskip

In summary, $\Ga_Z$ and $\Ga_{\rm inv}$ cannot exclude a massless
neutralino. The parts of the $\mu$-$M_2$~planes that lead to a large
deviation from the experimental values are mostly already excluded by
direct chargino searches.


\subsection{\boldmath{$W$}~Boson Mass \boldmath{$\MW$}, Effective
 Leptonic Mixing Angle \boldmath{$\sweff$} and \boldmath{$\Ga_Z$}}  
\label{subse:MWsweff}

Next we analyse the impact of light neutralinos on $\MW$ and $\sweff$
(and extend the $\Ga_Z$ analysis).  The most up-to-date calculations
for these quantities were presented in Refs.~\cite{MWweber} and
\cite{ZObsMSSM}. These predictions contain the full \order{\al} MSSM
result, as well as all known MSSM corrections beyond one-loop order,
in particular universal contributions entering via $\De\rho$ of
\order{\al \als}~\cite{dr2lA} and
\order{\al^2_{t,b}}~\cite{drMSSMal2A,drMSSMal2B}. Concerning
$\sweff$, in order to reproduce the best available SM
results~\cite{MWSM,dkappaSMbos2L} in the decoupling limit, also the
full electroweak~\order{\al^2} SM results
\cite{MWSMferm2L,MWSMbos2L,Hollik:2005va,BayernFermionic,UliMeierFussballgottBosonic},
mixed electroweak and QCD SM corrections
of~\order{\al\als}~\cite{drSMgfals}
and~\order{\al\als^2}~\cite{drSMgfals2}, as well as leading beyond
two-loop SM terms~\cite{drSMgf3mh0,drSMgf3,derhoalalscube} are
accounted for. We show these results as a function of $M_1$ in an
interval covering the solution of
\refeq{massless-neut-condition}, or equivalently given by
\begin{equation}
\label{M1mchismall}
M_1= 
\frac{M_2\,\MZ^2\sw^2\SZb}{M_2\,\mu - \MZ^2 \cw^2 \SZb}
+ \de M_1~,
\end{equation}
with $\de M_1$ ranging roughly from $-100$~GeV to $+100$~GeV.
Appropriate choices of $M_2$ and $\mu$ allow to analyse a $\neu{1}$
with a substantial bino, zino or higgsino component.

\medskip

In a first scenario a bino-like $\neu{1}$ is obtained by setting
$\tb=10$, $\msusy = 250\gev$, $A_{\tau}=A_t=A_b=\mu=\mgl=\MA=500\gev$,
$M_2=200\gev$ and $M_1$ according to \refeq{M1mchismall}, with the
above mentioned range for $\de M_1$.  This results in $\mneu{1}$
values between 0 and $\sim$$100\gev$.  What is of interest here is not
the absolute value of $\MW$ and $\sweff$ (and $\Ga_Z$), which again
strongly depends on the remaining SUSY parameters, but the change
induced by varying $M_1$ and thus also $\mneu{1}$. Clearly also these
effects are somewhat dependent on \textit{e.g.}  the sfermion mass
scale, but the effects due to the neutralino sector are the dominant
ones. The results are displayed in \reffi{fig:LightNeu1}, showing the
dependence of $\MW$ and $\sweff$ on $M_1$ and $\mneu{1}$. For
complementarity with \refse{subse:GammaZ}, we have included also the
result for $\Ga_Z$.  We show the two cases with $\Ga(Z \to
\neu{1}\neu{1})$ either included or not included into the total
$Z$~width.  The left column shows the results as a function of $M_1$,
while the right column as a function of the (corresponding) values of
$\mneu{1}$. The positive and negative values for $M_1$ result each in
one branch in the plots on the right.  As it happens for the chosen
set of parameters, $\Ga_Z$ and $\MW$ lie fully within the experimental
one sigma range, while $\sweff$ is outside the $1 \si$ boundaries
everywhere in the plot. As mentioned above, the relative agreement
between the MSSM predictions and the experimental results would have
been modified if we had chosen different $\msusy$, $A_{t,b}$, or $M_2$
(or other SUSY parameters).  As can be seen in \reffi{fig:LightNeu1},
none of the observables has a strong dependence on $M_1$. The shift
induced in $\MW$, for example, is $\approx 0.5\mev$ for the full $M_1$
range.  Even with an anticipated ILC precision of
$\de\MW^{\textup{ILC}}=7\mev$~\cite{mwgigaz,blueband} this is a
marginal effect. The situation is similar for $\sweff$. Here one can
observe a $\approx 0.4 \times 10^{-5}$ variation with $M_1$, which is
again marginal even compared with the ILC accuracy of
$\de\sweff^{\textup{ILC}}=1.3 \times 10^{-5}$~\cite{sw2gigaz}.  Also
the effects induced in $\Ga_Z$, even when including the $Z$-decay to
neutralinos, are insignificant compared to the experimental errors,
due to the small coupling of the bino-like $\neu{1}$ to the $Z$~boson.

\begin{figure}[htb!]
\begin{center}
\inclXMGdbleplot{figs/GammaZ-LightNeutralino-M1.eps}
\hspace{.3em}
\inclXMGdbleplot{figs/GammaZ-LightNeutralino-MNeu1.eps}\\[1em]
\inclXMGdbleplot{figs/MW-LightNeutralino-M1.eps}
\hspace{.3em}
\inclXMGdbleplot{figs/MW-LightNeutralino-MNeu1.eps}\\[1em]
\inclXMGdbleplot{figs/sweff-LightNeutralino-M1.eps}
\hspace{.3em}
\inclXMGdbleplot{figs/sweff-LightNeutralino-MNeu1.eps}
\end{center}
\vspace{-1em}
\caption{Total $Z$~width $\Ga_Z$ in- and excluding the process
  $Z\to\neu{1}\neu{1}$, $\MW$, and $\sweff$. The SUSY parameters are
  chosen as: $\tb=10$, $\msusy = 250\gev$, $A_{\tau}=A_t=A_b=\mu=\mgl
  =\MA=500\gev$, $M_2 = 200\gev$. $M_1$ is chosen according to
  \refeq{M1mchismall}, where $\de M_1$ is varied from $-100$ to
  $100\gev$, and the two branches in the plots on the right result
  from negative and positive $M_1$.  The green shading in the plots
  for $\Ga_Z$ and $\MW$ indicates that the whole area lies within the
  experimental $1 \sigma$ range.  }
\vspace{-1em}
\label{fig:LightNeu1} 
\end{figure}

\begin{figure}[htb!]
\begin{center}
\inclXMGdbleplot{figs/GammaZ-LightNeutralino2-M1.eps}
\hspace{.3em}
\inclXMGdbleplot{figs/GammaZ-LightNeutralino2-MNeu1.eps}\\[1em]
\inclXMGdbleplot{figs/MW-LightNeutralino2-M1.eps}
\hspace{.3em}
\inclXMGdbleplot{figs/MW-LightNeutralino2-MNeu1.eps}\\[1em]
\inclXMGdbleplot{figs/sweff-LightNeutralino2-M1.eps}
\hspace{.3em}
\inclXMGdbleplot{figs/sweff-LightNeutralino2-MNeu1.eps}
\end{center}
\vspace{-1em}
\caption{Total $Z$~width $\Ga_Z$ in- and excluding the process
  $Z\to\neu{1}\neu{1}$, $\MW$, and $\sweff$. The SUSY parameters are
  chosen to be $\tb=10$, $\msusy=A_{\tau}=A_t=A_b=\MA=\mgl=600\gev$,
  $\mu=125\gev$, $M_2=200\gev$. $M_1$ is set according to
  \refeq{M1mchismall}, where $\delta M_1=-100\gev$ to $+100\gev$, and
  the two branches in the plots on the right result from negative and
  positive $M_1$. The green shading indicates the experimental $1
  \sigma$ range of the three observables.  }
\vspace{-1em}
\label{fig:LightNeu12} 
\end{figure}

\medskip

As a second representative scenario, we study the case where $\tb =
10$, $\msusy = A_{\tau}=A_t=A_b=\MA=\mgl=600\gev$, $\mu=125\gev$, and
$M_2 =200\gev$. $M_1$ is set according to \refeq{M1mchismall}, where
$\de M_1=-100\gev$ to $+100\gev$. For this choice of parameters the
positive and negative values for $M_1$ result each in one branch in
the plots on the right, ranging up to $\mneu{1}\lsim70-90\gev$. The
smaller value of $\mu$ here allows for bigger non-bino like
$\neu{1}$-components. The numerical effects induced in the precision
observables are therefore expected to be bigger.  This expectation is
confirmed by~\reffi{fig:LightNeu12}, where as before the left column
shows the results as a function of $M_1$, while the right column as a
function of the (corresponding) values of $\mneu{1}$. The variation in
$\MW$ is $\approx 5\mev$, so roughly a factor ten bigger than before.
$\sweff$ shows a variation of $\approx$$\,2\times 10^{-5}$. Still,
even in comparison with the anticipated ILC precisions these are
barely observable effects. The largest effects in
\reffi{fig:LightNeu12} can be observed for the prediction of $\Ga_Z$.
The impact of the $Z$-boson decay into a pair of light neutralinos on
the total $Z$-boson width is seen to be quite substantial for this
choice of SUSY parameters. With today's experimental accuracy, a
$2\sigma$ effect is observable in the presence of a massless
neutralino.

\medskip

To conclude, even the most precise anticipated measurements of $\MW$
and $\sweff$ are not able to constrain the mass of $\neu{1}$. The only
possible constraints originate from the $Z$~decay width, although even
these are not very powerful for most of the parameter space, as they
require a rather sizable coupling of $\neu{1}$ to the $Z$~boson. Most
of the parameter space which fulfills this requirement is already
ruled out by direct chargino searches, as discussed in
\refse{subse:GammaZ}. The constraints are also easily 
satisfied by increasing the bino component of the light neutralino.


\subsection{Electric Dipole Moments and Anomalous Magnetic Moment of
the Muon}
\label{subse:EDMgm2}

The effect of a small or vanishing mass of the lightest neutralino on
electric dipole moments (EDMs) and on the anomalous magnetic moment of
the muon, $(g-2)_\mu$, is shown in \reffi{fig:edmgm2}.  For the EDMs
the one- and two-loop formulas of
\citeres{EDMPilaftsis,EDMRitz,EDMheavy,EDMmiracle} have been used, see
\citeres{EDMrev1,EDMrev2} for reviews. The SUSY contributions to
$(g-2)_\mu \equiv 2 \amu$ are based on the one- and two-loop formulas
given in \citeres{g-2MSSMf1l,g-2MSSMlog2l,g-2FSf,g-2CNH}, see
\citere{g-2reviewDS} for a recent review. These calculations have been
performed using the computer code {\tt FeynHiggs}
\cite{feynhiggs,mhiggslong,mhiggsAEC,mhcMSSMlong}.

\medskip

For the various EDM measurements upper limits exist as given in
\refta{tab:edm}. For the anomalous magnetic moment of the muon a
``discrepancy'' of the experimental result from the SM prediction has
been observed~\cite{DDDD}
\begin{align}
\amuexp-\amutheo &= (27.5 \pm 8.4) \times 10^{-10},
\label{delamu}
\end{align}
equivalent to a 3.3$\,\sigma$ effect.%
\footnote{Three other recent evaluations yield slightly different
  numbers~\cite{g-2reviewFJ,g-2HMNT2,g-2reviewMRR}, 
  but similar discrepancies with the SM prediction.}%
~While SUSY contributions could easily explain this ``discrepancy'', a
massless neutralino could in principle lead to a too large contribution
to $\amu$.

\begin{table}[htb!]
\renewcommand{\arraystretch}{1.2}
\begin{center}
\begin{tabular}{|c|c|c|} \hline
System & limit & group \\ \hline\hline
$e^-$  & $1.6 \times 10^{-27}$ (90\% C.L.) & Berkeley \\ \hline
$n$    & $2.9 \times 10^{-26}$ (90\% C.L.) & ILL     \\ \hline
$\mbox{}^{199}$Hg & $2.1 \times 10^{-28}$  (95\% C.L.) & Seattle \\ 
\hline \hline
\end{tabular}
\caption{Present bounds for EDMs~\cite{EDMe,EDMn,EDMhg}.}
\label{tab:edm}
\end{center}
\vspace{-1em}
\end{table}

\begin{figure}[h!]
\begin{center}
\includegraphics[width=7cm,height=6cm]{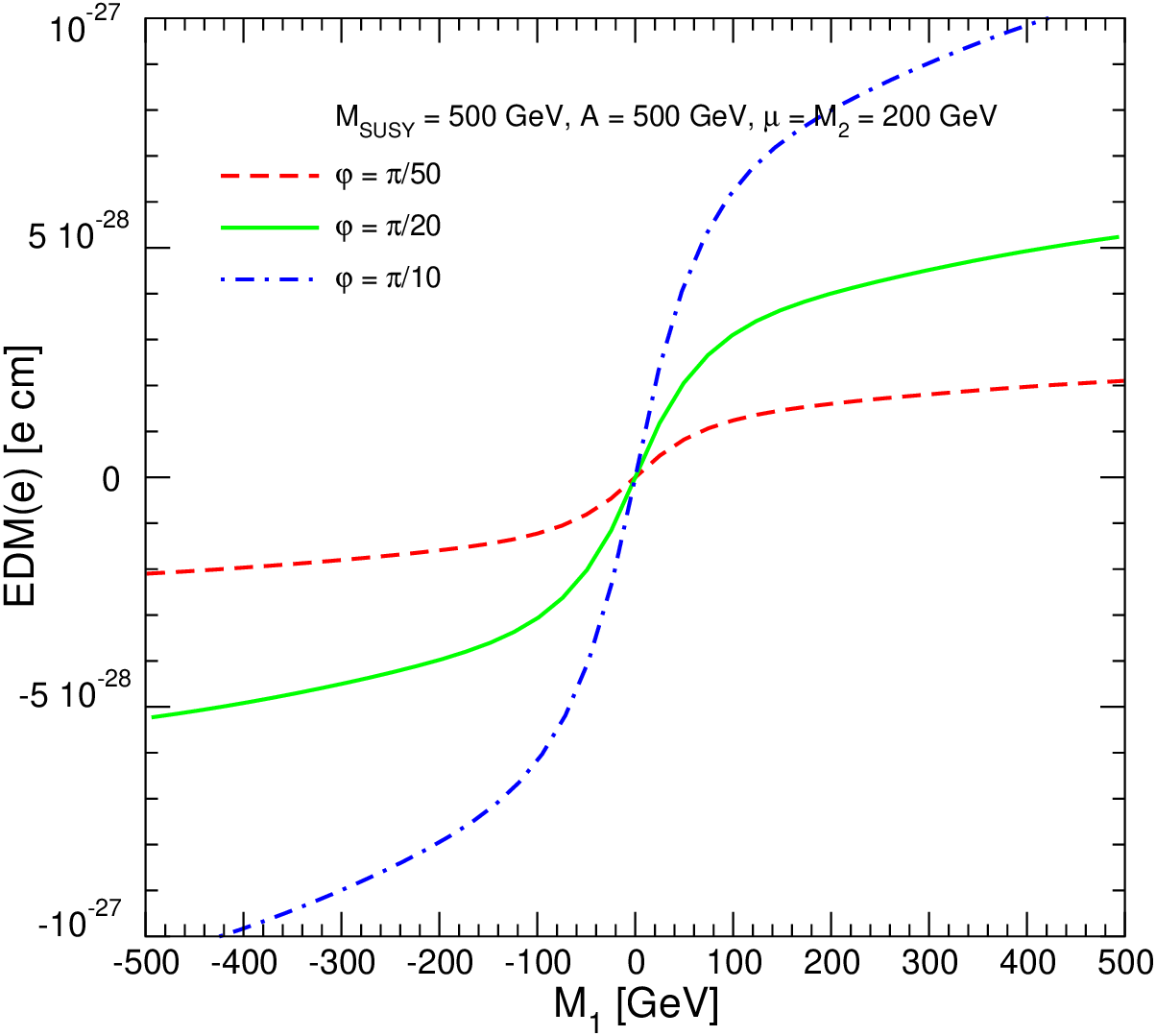}
\hspace{.3em}
\includegraphics[width=7cm,height=6cm]{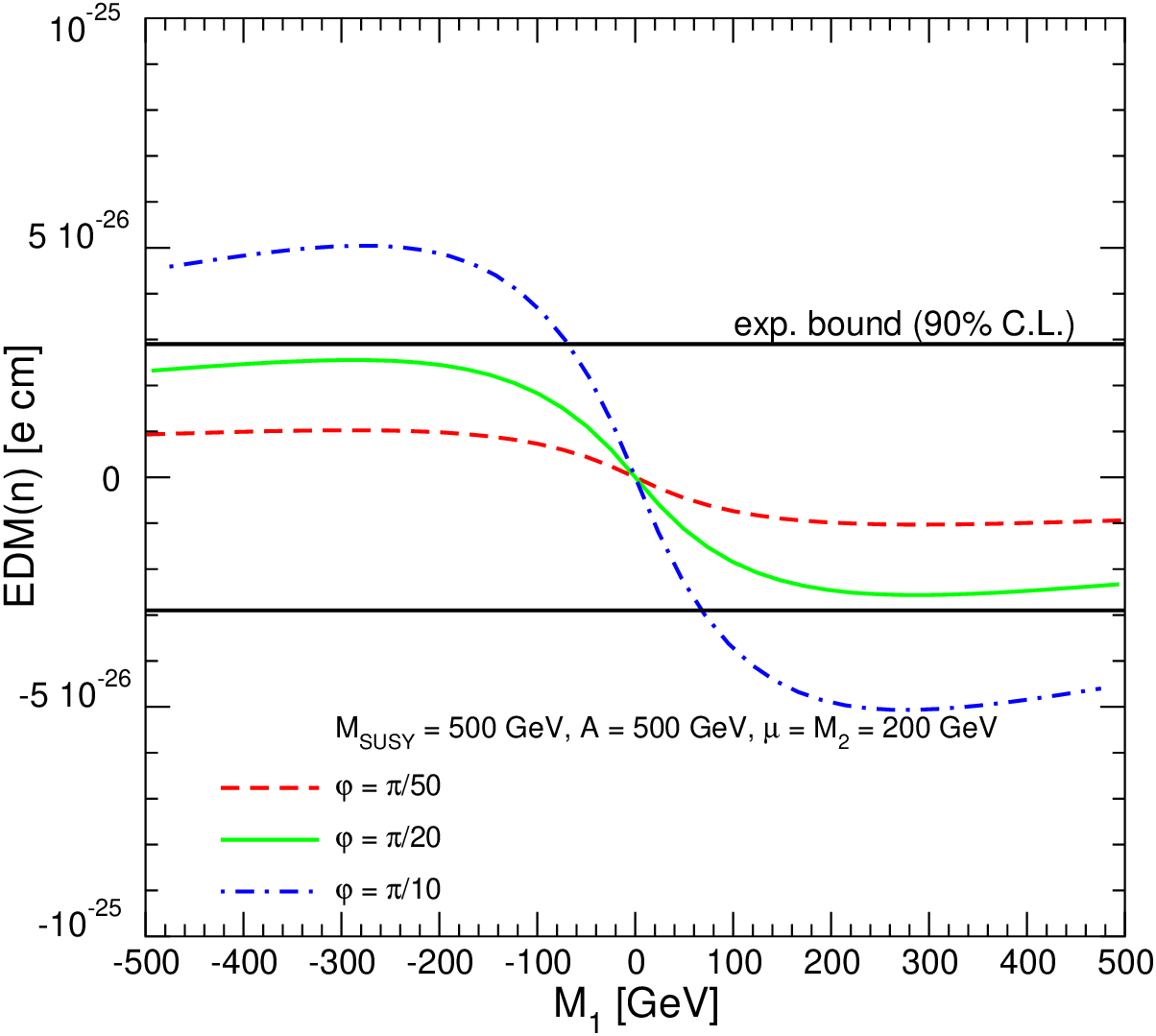}\\[.3em]
\includegraphics[width=7cm,height=6cm]{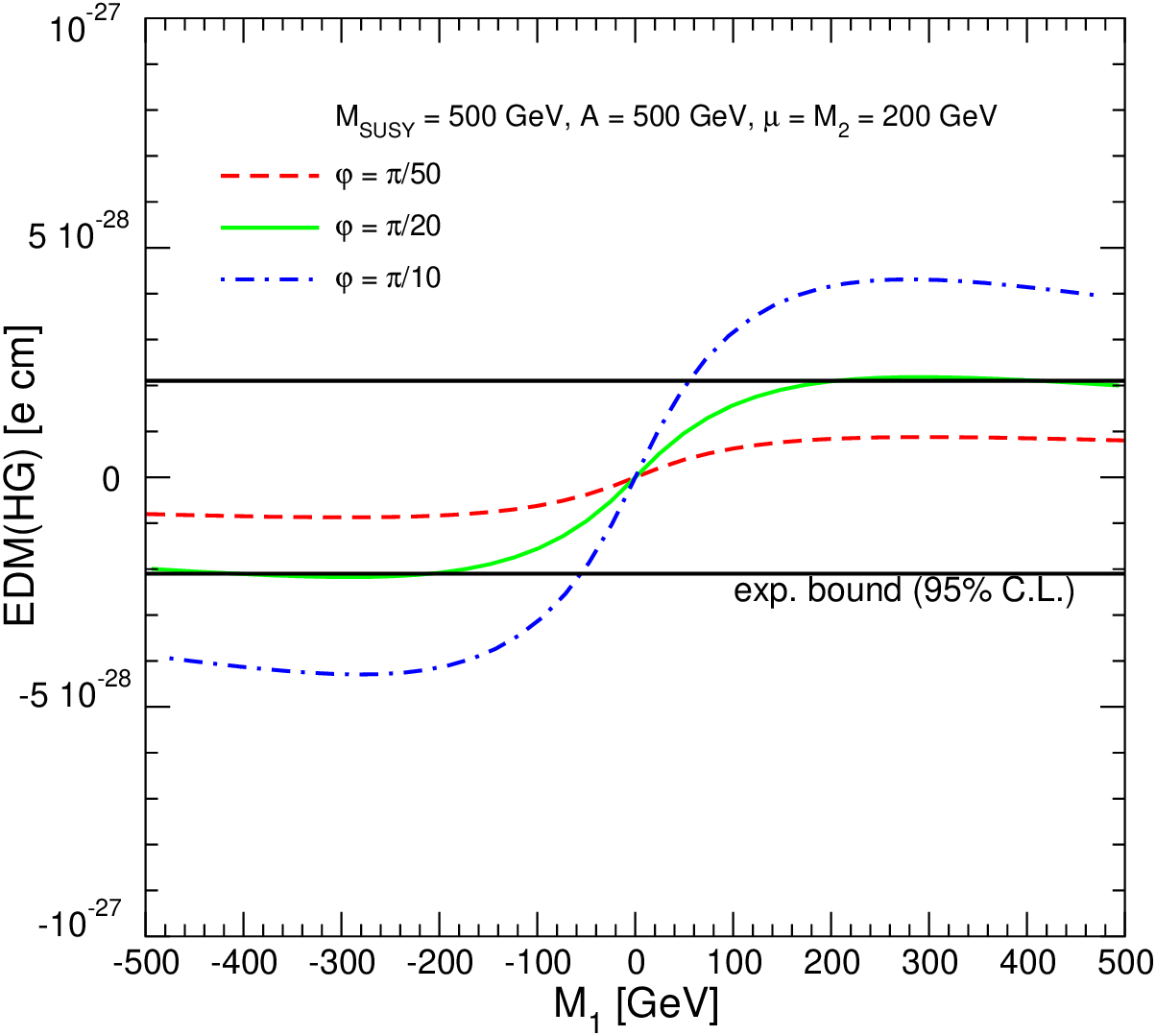}
\hspace{.3em}
\includegraphics[width=7cm,height=6cm]{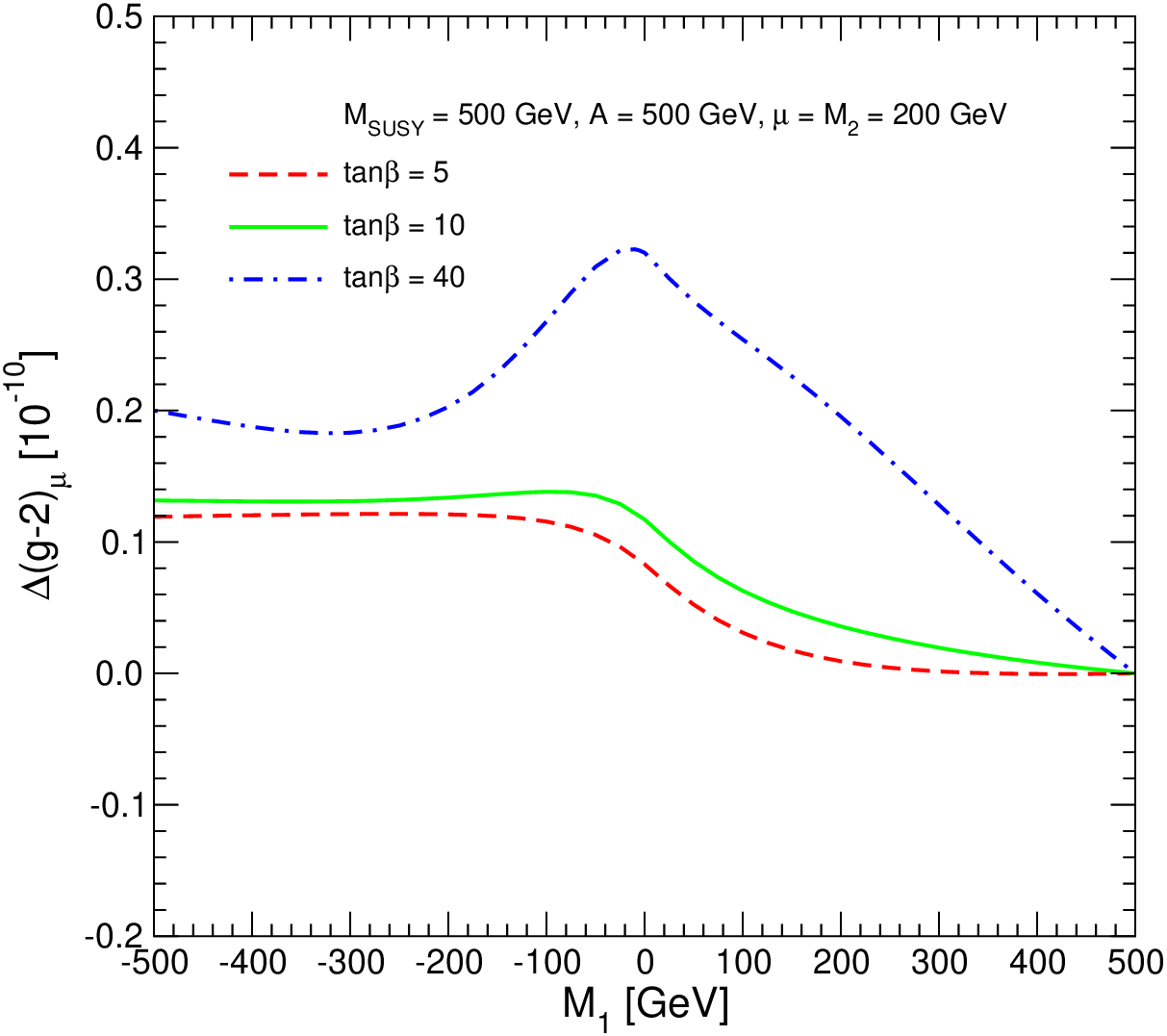}
\end{center}
\vspace{-2.0em}
\caption{The EDMs of the electron (upper left plot), the neutron (upper
  right) and mercury (lower left), as well as the shift in $(g-2)_\mu$
  normalised to $(g-2)_\mu(M_1 = 500 \gev)$ (lower right plot), are
  shown as a function of $M_1$ (see text). The other parameters are
  $\MHp = \msusy = 500 \gev$, $A_f = 500 \gev$, $M_2 = \mu = 200
  \gev$. For the EDMs we have fixed $\tb = 10$ and set the phase of
  $M_1$ to $\phiMe = \pi/50, \pi/20, \pi/10$.  For the shift in
  $(g-2)_\mu$ we fixed the phases to be zero, but varied $\tb = 5, 10,
  40$. $\mneu{1} = 0$ is reached around $M_1 \approx 0$.}
\label{fig:edmgm2} 
\vspace{-0.5em}
\end{figure}

The SUSY contribution to the EDMs and to $(g-2)_\mu$, shown in
\reffi{fig:edmgm2}, have been evaluated as a function of $M_1$ with
the other parameters set to $\msusy = \MHp = 500 \gev$, $A_f=500\gev$,
$M_ 2=\mu=200\gev$. For the EDMs we have fixed $\tb=10$ and set the
phase of $M_1$ to $\phiMe = \pi/50, \pi/20, \pi/10$.  For $(g-2)_\mu$,
we fixed the phases to be zero, but varied $\tb=5,10,40$.  According
to Eqs.~(\ref{massless-neut-condition}), (\ref{massless-neut-value}),
(\ref{eq:CPcase1}), (\ref{eq:CPcase2}) a massless neutralino is
reached around $M_1 \approx 0$. We show the contribution relative to
$(g-2)_\mu(M_1 = 500 \gev)$, which is \order{10^{-9}} for this set of
parameters.

The SUSY contributions to the EDMS involving the lightest neutralino
go to zero for vanishing $\mneu{1}$, as can be seen in the upper left,
upper right and lower left plot in \reffi{fig:edmgm2}, where we show
the EDM of the electron, the neutron and mercury, respectively.
Consequently no lower bound on $\mneu{1}$ can be set. For large values
of $\phiMe$, on the other hand, an upper limit on $M_1$ and thus on
$\mneu{1}$ can be derived as can be seen in the upper right and lower
left plot of \reffi{fig:edmgm2}. The variation of $(g-2)_\mu$, shown
in the lower right plot of \reffi{fig:edmgm2}, stays below $\sim 0.5
\times 10^{-10}$ and is thus well below the current uncertainty of
$8.4 \times 10^{-10}$~\cite{DDDD}. Therefore, while (for this exemplary 
set of parameters) the SUSY contributions is of the same size as the 
deviation between the SM and the experimental result, no experimental 
limit on $\mneu{1}$ can be set.

}

}

\section{Rare Meson Decays}
\label{meson-decays}
{
\renewcommand{\bino}{\widetilde B}

In the SM, $^3\!S_1$ mesons (vector mesons) can decay into
a neutrino pair,
\textit{e.g.} 
\begin{equation}
J/\psi(1S),\,\Upsilon(1S)\rightarrow \nu_i \,
\bar\nu_i,\,\qquad\qquad i=e,\mu,\tau\,,
\label{inv1}
\end{equation}
where $J/\psi$ and $\Upsilon$ denote $c\bar{c}$ and $b\bar{b}$ (ground)
states, respectively. 
Below we also investigate $\phi= s\bar{s}$ and 
the light mesons $\rho$ and $\omega$, which are
superpositions of $u \bar{u}$ and $d \bar{d}$ states. 
An example Feynman graph for a meson decay into a neutrino pair
at the parton level is shown in Fig~\ref{subfig:ccdecay}.
If we allow for neutrino masses, which enable a chirality flip, also
$^1\!S_0$ mesons (pseudoscalar mesons) can decay into a neutrino pair, 
\textit{e.g.}
\cite{Kayser:1974ct,Herczeg:1981xa}
\begin{equation}
\pi^0\rightarrow \nu_i \,\bar\nu_i\,\qquad\qquad i=e,\mu,\tau\,,
\label{inv2}
\end{equation}
where again Fig.~\ref{subfig:ccdecay} represents the tree-level graph.
In both cases, the neutrino pair remains unobservable. 
It is possible to look for the invisible decay of 
mesons, Eqs.~(\ref{inv1}), (\ref{inv2}), for example via the decay chain
\begin{eqnarray}
\psi(2S)&\rightarrow& J/\psi+\pi\pi\qquad\qquad i=e,\mu,\tau\,, \label{hook}\\ 
&& \hookrightarrow \nu_i\bar\nu_i \nonumber
\end{eqnarray}
by tagging on the invariant mass of the di-pion system. To our
knowledge, this idea was first proposed in Ref.~\cite{Rich:1976nu}, as
a test of the SM neutral current, and has since been widely employed
\cite{Besson:1984ha,Balest:1994ch,Ablikim:2006eg,Ablikim:2007ek}. In
Ref.~\cite{Fayet:1979qi} it was proposed as a method to look for
physics beyond the SM, then a light gravitino. It has since been used
to look for many different aspects of beyond the SM physics
\cite{Campbell:1983jx,Fayet:1991ux,Adhikari:1994wh,Adhikari:1994iv,Adhikari:1995bq,Chang:1997tq,McElrath:2005bp,Fayet:2006sp}.

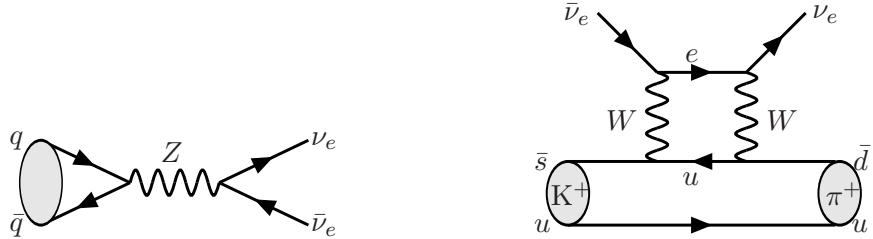
\begin{figure}[h!]
\centering
\subfigure[Meson decay into a neutrino pair.\label{subfig:ccdecay}]{
\unitlength=1.6 pt
\SetScale{1.6}
\SetWidth{0.7}      
\begin{picture}(95,50)(0,0)
\Text(12.0,40.0)[r]{$q$}
\ArrowLine(16.0,40.0)(37.0,30.0) 
\Text(12.0,20.0)[r]{$\bar{q}$}
\ArrowLine(37.0,30.0)(16.0,20.0) 
\Text(47.0,35.0)[b]{$Z$}
\Photon(37.0,30.0)(58.0,30.0){3.0}{4} 
\Text(80.0,40.0)[l]{$\nu_e$}
\ArrowLine(58.0,30.0)(79.0,40.0) 
\Text(80.0,20.0)[l]{$\bar{\nu}_e$}
\ArrowLine(79.0,20.0)(58.0,30.0)
\GOval(16,30)(10,5)(0){0.9}
\end{picture} \
}
\subfigure[$\mathrm{K}^+$ decay into a neutrino pair and a pion
      \label{subfig:kaondecay}]{
\unitlength=1.6 pt
\SetScale{1.6}
\SetWidth{0.7}      
{} \qquad\allowbreak
\begin{picture}(95,75)(0,0)
\Text(8,20)[l]{$u$}
\Text(87,20)[r]{$u$}
\ArrowLine(16,20)(79,20)
\Text(8,35)[l]{$\bar{s}$}
\Text(87,36)[r]{$\bar{d}$}
\Text(47,31)[r]{$u$}
\ArrowLine(79,35)(16,35)
\Text(25,45)[l]{$W$}
\Photon(37,35)(37,56){2.5}{4}
\Text(63,45)[l]{$W$}
\Photon(58,35)(58,56){2.5}{4}
\Text(47,60)[r]{$e$}
\ArrowLine(37,56)(58,56)
\Text(15,70)[l]{$\bar{\nu}_e$}
\ArrowLine(23,70)(37,56)
\Text(80,70)[r]{$\nu_e$}
\ArrowLine(58,56)(72,70)
\GOval(16,27.5)(7.5,5)(0){0.9}
\Text(12,27.5)[l]{$\mathrm{K}^+$}
\GOval(80,27.5)(7.5,5)(0){0.9}
\Text(77,27.5)[l]{$\pi^+$}
\end{picture} \
}
\caption{Example diagrams for meson decays into a neutrino pair}
\label{fig:mesontonp}
\end{figure}

Mesons can also decay to a lighter meson and a neutrino pair,
\textit{e.g.} \cite{Adler:2001xv,Adler:2008zz}
\begin{equation}
K^+\rightarrow\pi^+\nu_i\bar\nu_i,\qquad\qquad i=e,\mu,\tau\,.
\label{kpi}
\end{equation}
A parton level Feynman graph is shown in Fig.~\ref{subfig:kaondecay}.
The rare decay, Eq.~(\ref{kpi}), has a similar signature to the cascade,
Eq.~(\ref{hook}), and has in fact been observed
\cite{Adler:2001xv,Adler:2008zz}. The partial decay width is
consistent with the SM \cite{Inami:1980fz,Buchalla:1993bv}, albeit
with a still large experimental error. It has also been used to
investigate beyond the SM physics
\cite{Singh:1970aq,Gaillard:1982rw,Ellis:1982ve,Keung:1983hx,Nieves:1985ir,Buras:1997ij,Deshpande:2004xc}.

In the P$_{\!6}$ conserving MSSM with a stable, light neutralino, we
can replace the neutrino pair in the decays Eqs.~(\ref{inv1}),
(\ref{inv2}) and (\ref{kpi}) by a pair of light neutralinos. The
latter would also remain unobservable and we would obtain
similar signatures, possibly slightly modified by the light neutralino
mass. We thus have potential new decay modes which can be used to
search for, or constrain, the light neutralino scenario
\cite{Gaillard:1982rw,Ellis:1982ve,Keung:1983hx,Campbell:1983jx,Haber:1984rc,Dobroliubov:1987cb,McElrath:2005bp,Adhikari:1994iv,Adhikari:1994wh,Adhikari:1995bq}.
The mass range of the decaying mesons (\textit{e.g.} $m_\pi \approx
130\MeV$, $m_K \approx 500\MeV$, $m_\Upsilon \approx 10\GeV$) enables
a test of various light neutralino masses \cite{Ellis:1982ve}.

Motivated by our discussion in Sects.~\ref{framework} and
~\ref{prec-observ}, we assume throughout this section that the
neutralino is pure bino, denoted as $\bino$, and very light,
\textit{i.e.} that the neutralino decay mode is kinematically
accessible to the meson. We recall that a pure bino has no interaction
with the $Z$ boson and its couplings to right-handed sfermions are
stronger than to left-handed sfermions. The corresponding
supersymmetric Feynman diagrams for the meson decays can be quite
different than those for the SM decays. We list a few examples in
Fig.~\ref{fig:mesontobino}.

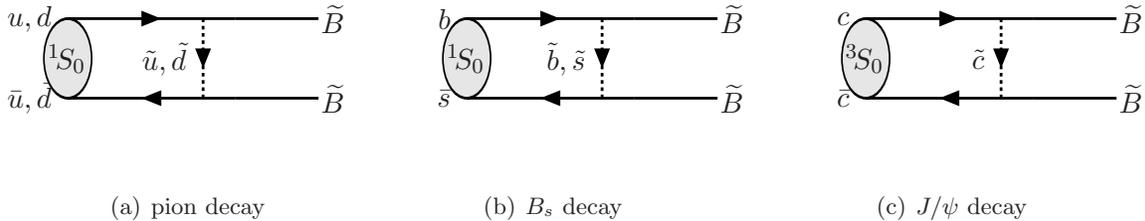
\begin{figure}[h!]
\centering
\subfigure[pion decay\label{subfig:pitobino}]
{
\unitlength=1.5 pt
\SetScale{1.5}
\SetWidth{0.7}      
\begin{picture}(95,50)(0,0)
\Text(12.0,40.0)[r]{$u,d$}
\ArrowLine(16.0,40.0)(58.0,40.0) 
\Text(80.0,40.0)[l]{$\bino$}
\Line(58.0,40.0)(79.0,40.0)
\Text(46.0,30.0)[r]{$\tilde{u},\tilde{d}$}
\DashArrowLine(50.0,40.0)(50.0,20.0){1.0} 
\Text(12.0,20.0)[r]{$\bar{u},\bar{d}$}
\ArrowLine(58.0,20.0)(16.0,20.0) 
\Text(80.0,20.0)[l]{$\bino$}
\Line(58.0,20.0)(79.0,20.0)
\GOval(16,30)(10,6)(0){0.9}
\Text(11,30)[l]{${}^1\!S_0$}
\end{picture} \ 
}
\hspace*{-15mm}
\subfigure[$B_s$ decay\label{subfig:bstobino}]
{
\unitlength=1.5 pt
\SetScale{1.5}
\SetWidth{0.7}      
{} \qquad\allowbreak
\begin{picture}(95,50)(0,0)
\Text(12.0,40.0)[r]{$b$}
\ArrowLine(16.0,40.0)(58.0,40.0) 
\Text(80.0,40.0)[l]{$\bino$}
\Line(58.0,40.0)(79.0,40.0)
\Text(46.0,30.0)[r]{$\tilde{b},\tilde{s}$}
\DashArrowLine(50.0,40.0)(50.0,20.0){1.0} 
\Text(12.0,20.0)[r]{$\bar{s}$}
\ArrowLine(58.0,20.0)(16.0,20.0) 
\Text(80.0,20.0)[l]{$\bino$}
\Line(58.0,20.0)(79.0,20.0)
\GOval(16,30)(10,6)(0){0.9}
\Text(11,30)[l]{${}^1\!S_0$}
\end{picture} \ 
}
\hspace*{-15mm}
\subfigure[$J/\psi$ decay\label{subfig:jpsitobino}]
{
\unitlength=1.5 pt
\SetScale{1.5}
\SetWidth{0.7}      
{} \qquad\allowbreak
\begin{picture}(95,50)(0,0)
\Text(12.0,40.0)[r]{$c$}
\ArrowLine(16.0,40.0)(58.0,40.0) 
\Text(80.0,40.0)[l]{$\bino$}
\Line(58.0,40.0)(79.0,40.0)
\Text(46.0,30.0)[r]{$\tilde{c}$}
\DashArrowLine(50.0,40.0)(50.0,20.0){1.0} 
\Text(12.0,20.0)[r]{$\bar{c}$}
\ArrowLine(58.0,20.0)(16.0,20.0) 
\Text(80.0,20.0)[l]{$\bino$}
\Line(58.0,20.0)(79.0,20.0)
\GOval(16,30)(10,6)(0){0.9}
\Text(11,30)[l]{${}^3\!S_0$}
\end{picture} \ 
}
\caption{Example diagrams for the decay of a $\pi$, $B_s$, and $J/\psi$-meson
            into a bino pair.}
\label{fig:mesontobino}
\end{figure}

In the following, we give an overview over the present bounds existing
in the literature, modifying them to binos, where necessary. The
analyses carried out so far involve significant approximations. An
up-to-date, detailed analysis of meson decays to light neutralinos, in
particular including the necessary higher-order calculations, is
beyond the scope of this paper and will be presented in
Ref.~\cite{work-ip:2008}.  However, given the current status of the
data, \textit{cf.}  Table~\ref{tab:meson}, we expect the present
conclusions to be robust.


\subsection{Pseudoscalar Mesons}
Pseudoscalar mesons (psm), \textit{e.g.} $\pi^0$, $K$, $\eta$,
$\eta^\prime$, and $B_s$ can in principle decay into a light
neutralino pair. The related photino decay $\pi^0\rightarrow
\photino\photino$ was first computed in
Refs.~\cite{Gaillard:1982rw,Haber:1984rc}. Later, in
Ref.~\cite{Dobroliubov:1987cb}, the decays $\pi^0\to\photino\photino$
and $\pi^0\to\photino\photino \gamma$ were investigated. If we rescale
the result in Ref.~\cite{Dobroliubov:1987cb} from a squark mass of
$m_{\tilde q}=70\,\mathrm{GeV}$ to a squark mass of $m_{\tilde
q}=300\,\mathrm{GeV}$ \cite{Abazov:2007ww}, we obtain an upper bound
on the photino branching ratios, which is well below the
current experimental limits for the neutrino
decays~\cite{Amsler:2008zz}
\begin{eqnarray}
\mathrm{BR}(\pi^0\rightarrow\nu\bar{\nu})&\lsim& 2.7\cdot 10^{-7}\,,\\
\mathrm{BR}(\pi^0\rightarrow\nu\bar{\nu}\gamma)&\lsim& 6\cdot10^{-4}\,.
\end{eqnarray}
This conclusion also holds for the corresponding bino decays. For
their theoretical estimates the authors in
Ref.~\cite{Dobroliubov:1987cb} assumed a large left--right mixing in
the squark sector, not taking into account flavour changing neutral
current constraints. Due to the very
small supersymmetric branching ratio, no bound is obtained on the bino
mass.

In Ref.~\cite{Gaillard:1982rw,Haber:1984rc} no left--right mixing was
assumed. We reestimate their branching ratio for a pure bino and for
heavier squarks and assume that the relevant squarks are degenerate.
We also generalize to an arbitrary psm. The corresponding Feynman
graph for a pion is shown in Fig.~\ref{subfig:pitobino}.  The decay
width to a bino pair is then given by
\begin{equation}
\label{mesondecay}
\Gamma(\mathrm{psm}\to\tilde{B}\tilde{B}) =
                  \frac{\pi \alpha^2 }{4\cos^4\theta_{\mathrm{w}}}C(\mathrm{psm})
                 \frac{m_{\bino}^2 m_{\mathrm{psm}}}{m_{\tilde{q}}^4}
                 \sqrt{1 - \dfrac{4 m_{\bino}^2}{m_{\mathrm{psm}}^2}} \enspace .
\end{equation}
The constant $C$ depends on the considered meson and is given by
\begin{align}
C(\pi^0) &\,=\, \frac{2}{9} f^2_{\pi^0}\enspace,\\[1mm]
C(\eta)  &\,=\, \frac{3}{4}\Bigl(\sin\theta_{1} f_1 -
                  \tfrac{2\sqrt{2}}{9}\cos\theta_{8}f_8\Bigr)^2\enspace,\\[1mm]
C(\eta^\prime)  &\,=\, \frac{3}{4}\Bigl(\sin\theta_{1} f_1 +
                  \tfrac{2\sqrt{2}}{9}\cos\theta_{8}f_8\Bigr)^2\enspace,\\[1mm]
C(B_s) &\,=\, \frac{|V_{cb}|^2 c^2 f_{B_{s}}^2}{216\pi^2}\enspace,
\end{align}
where $f_{\pi^0}$, $f_1$, $f_8$, and are the decay constant of the
$\pi^0$, $\eta_1$, $\eta_8$, and $B_s$, respectively; $\theta_{1,8}$
are the mixing angles between the $\eta_1$ and $\eta_8$ states.
$V_{cb}$ is the charm--bottom CKM matrix element. These are all given
in \citeres{Yao:2006px,Amsler:2008zz}. $c$ parameterises the flavour
mixing in the squark sector; representative values for $c$ and
$f_{B_s}$ are taken from Ref.~\cite{Adhikari:1994wh}.

At the parton level, the decay $B_s \to \bino\bino$ proceeds via $b
\to s\bino\bino$, \textit{cf.} Fig.~\ref{subfig:bstobino}. The flavour
changing neutral current (FCNC) is possible because the left-handed
squark mass matrices and the quark mass matrices are not (necessarily)
simultaneously diagonal~\cite{Adhikari:1994wh}. This mismatch is
parameterised by the constant $c$. The authors of
Ref.~\cite{Adhikari:1994wh} obtain for a $80\GeV$ squark mass a
branching ratio of about $\mathcal{O}(10^{-7} - 10^{-5})$, depending
on the amount of flavour violation. In Table~\ref{tab:meson}, we have
rescaled their result and present the branching ratio for a squark
mass of $300\GeV$.

In \refta{tab:meson}, we list in the upper part the numerical values
for the branching ratios for the various meson decays to binos. Here
we have assumed $m_{\tilde q}=300\,\mathrm {GeV}$. Note that the decay
width, Eq.~(\ref{mesondecay}), peaks at a value $m_{\bino}=m_{\mathrm
{psm}}/\sqrt{6}$. In order to obtain conservative bounds in
Table~\ref{tab:meson}, we assume in turn for each meson this value for
the bino mass. It can be seen that the branching ratios for a meson
decay into a light bino pair are several orders of magnitude lower
than the experimental bounds on the respective invisible widths.  This
is mainly due to the small $\bino$ mass in the numerator (because of
the helicity flip required for a decay of a pseudoscalar into a pair
of binos) and the large squark masses in the denominator of
\refeq{mesondecay}, suppressing this weak decay mode.  Consequently no
bounds on a light $\bino$ can be set from these decay modes.

\begin{table}[t!]
\renewcommand{\arraystretch}{1.4}
\centering
\begin{tabular}{c|ccc}
\hline\hline
decay & maximal BR & experimental bound on BR & literature \\
\hline\hline
$\pi^0 \to \bino\bino$ & $\mathcal{O}(10^{-13})$ & $ < 2.7\cdot 10^{-7}$ &
\cite{Herczeg:1981xa,Dobroliubov:1987cb,Artamonov:2005cu} \\
$\eta \to \bino\bino$  & $\mathcal{O}(10^{-13})$ & $< 6\cdot 10^{-4}$&
\cite{Ablikim:2006eg} \\
$\eta^\prime\to\bino\bino$ & $\mathcal{O}(10^{-14})$ & $< 1.4 \cdot 10^{-3}$&
\cite{Ablikim:2006eg} \\
$B_s \to\bino\bino$ & $\mathcal{O}(10^{-8})$ & --- &
\cite{Adhikari:1994wh,Adhikari:1994iv,Adhikari:1995bq}\\
\hline
$\phi\to\bino\bino$ & $\mathcal{O}(10^{-16})$ & --- &
\cite{Keung:1983hx}\\
$J/\psi\to \bino\bino$ & $\mathcal{O}(10^{-11})$ & $< 7.1\cdot 10^{-4}$ &
\cite{Keung:1983hx,Ablikim:2007ek}\\
$\Upsilon(1S)\to\bino\bino$ & $\mathcal{O}(10^{-10})$ & $< 2.5 \cdot 10^{-3}$&
\cite{Ellis:1983ed,Keung:1983hx,Tajima:2006nc,Rubin:2006gc}\\
$\Upsilon(1S)\to \mathrm{inv.}$ &  & $< 1.07 \cdot 10^{-3}$ &
\cite{McElrath:2005bp} \\
$\rho\to\bino\bino$ & $\mathcal{O}(10^{-15})$ & --- & \\
$\omega\to\bino\bino$ & $\mathcal{O}(10^{-16})$ & --- & \\
\hline
$K^+ \to \pi^+ \bino\bino$ &$\mathcal{O}(10^{-{15}})$ & $1.5^{+1.3}_{-0.9} \times 10^{-10}$ &
\cite{Gaillard:1982rw,Ellis:1982ve}\\
\hline\hline
\end{tabular}
\caption{
  Comparison of maximum values of various branching ratios of meson 
  decays to light binos. We show the theoretical maximum value and the
  current experimental bounds on the BR. For the theoretical 
  computation we choose the bino mass to maximize the BR. The upper 
  (middle) part shows the results for the pseudoscalar (vector) mesons,
  the lower part shows a rare decay. For the pseudoscalar mesons, the 
  squark masses are set to $300 \GeV$. For the vector mesons, see 
  \refeqs{eq:crho}--(\ref{eq:cupsilon}). All data are taken from 
  \citeres{Yao:2006px,Amsler:2008zz} and from the specified
  literature. 
  ``---'' indicates that no experimental bound has been derived yet. 
   We have included also for 
  comparison a related experimental result on the invisible decay of
  the $\Upsilon$ \cite{McElrath:2005bp}.}
\label{tab:meson}
\renewcommand{\arraystretch}{1.0}
\end{table}


\subsection{Vector Mesons and Quarkonium Decay}

The decays of vector mesons to photinos were first considered in
Refs.~\cite{Ellis:1983ed,Campbell:1983jx,Haber:1984rc}. The decay of
the various $\Upsilon(n S)$ (excited) states into a neutralino pair
and its discovery possibility at $B$~factories is discussed in detail
in Ref.~\cite{McElrath:2005bp}.

In \citere{Ellis:1983ed}, the decay of quarkonium into gluinos and
photinos has been calculated.  Theses decays violate parity.  The
coupling strength of photinos and gluinos to left- and right-handed
squarks are equal. Consequently, the masses of left-handed and
right-handed squarks must be different to allow for these decays.
However, the first argument does \emph{not} hold for the decay into
binos: The coupling strength to right-handed particles is larger than
to left-handed particles, \textit{i.e.} parity is explicitly violated.
Hence, this decay is also possible when left- and right-handed squark
masses are equal to each other.

The quarkonium decay width can be approximated by, \textit{cf.} 
\citere{Barger:1987nn},
\begin{equation}
\Gamma(^3\!S_1(q\bar{q}) \to \tilde{B}\tilde{B}) =
      \frac{C_V}{\cos^4\theta_{\mathrm{w}}} \Gamma(V\to e^+ e^-)
      \frac{m_V^4}{m_{\tilde{q}}^4}
      \biggl(1 - \frac{4 m_{\bino}^2}{M^2_{V}}\biggr)^{3/2} ,
\end{equation}
where $m_V$ is the mass of the vector meson $V =\,^3\!S_1(q\bar{q}$),
and $m_{\tilde{q}}$ is the mass of the exchanged squark.  $C_V$ is a
constant depending on the sum of the squared
quark--squark--Bino--coupling strengths.  For the different mesons,
$C_V$ is given by
\begin{align}
\label{eq:crho}
C_\rho & = \frac{289}{10368},\quad  m_{\tilde{u}_L} = m_{\tilde{d}_R}\to \infty,\;\;
                  m_{\tilde{u}_R} = m_{\tilde{d}_L} \equiv m_{\tilde{q}} 
                   =300\GeV\\[2mm]
C_\omega &= \frac{25}{72},  \quad \,\;\;\;\;\;m_{\tilde{u}_L} = m_{\tilde{d}_L}\to 
\infty, \;\;
                  m_{\tilde{u}_R} = m_{\tilde{d}_R} \equiv m_{\tilde{q}} 
                   =300\GeV\\[2mm]
C_{\phi} &=  \frac{1}{72},\qquad \qquad \qquad m_{\tilde{s}_L}\to \infty, \;\;\;\;
\qquad\quad
      m_{\tilde{s}_R} \equiv m_{\tilde{q}} = 300\GeV\\[2mm]
C_{J/\psi} & = \frac{1}{18},\qquad \qquad \qquad  m_{\tilde{c}_2}\to \infty, \;\qquad\qquad
       m_{\tilde{c}_1} \equiv m_{\tilde{q}} = 300\GeV,
      \quad \theta_{\tilde c} = \frac{\pi}{2}\\[2mm] \label{eq:cupsilon}
C_{\Upsilon} &=\frac{1}{72},\qquad \qquad \qquad  m_{\tilde{b}_2}\to \infty,\;\qquad\qquad
      m_{\tilde{c}_1} \equiv m_{\tilde{q}} = 300\GeV,
      \quad \theta_{\tilde b} = \frac{\pi}{2}\,,
\end{align}
where we indicate the approximations we have made for the squark
masses. In \refeqs{eq:crho}--(\ref{eq:cupsilon}), we have also
listed the input parameters (squark masses and mixing angles) employed
to derive the upper bounds on the branching ratio. They are chosen in
such a way that the branching ratio is maximal for a massless
neutralino. For the $u$, $d$, and $s$ squark, we assume that left- and
right-handed squarks do not mix and that all squark masses are
degenerate.  For the $c$ and $b$ squark, we include squark mixing.
$\theta_{\tilde{c}}$ and $\theta_{\tilde{b}}$ denote the mixing angles
in the scalar charm and bottom sector, respectively.

As in the pseudoscalar meson case, the decays are strongly suppressed
by the large squark masses.  The results of the theory evaluation are
compared to the experimental data in the middle part of
\refta{tab:meson}. It can be seen that no bounds on the mass of a
light bino can be deduced. The discrepancy between the experimental
bound and the theoretical estimate is quite large. We thus expect
these results to be quite robust. More detailed formul{\ae} can be
found in \citere{work-ip:2008}.


\subsection{Loop-induced Meson Decays to Binos}

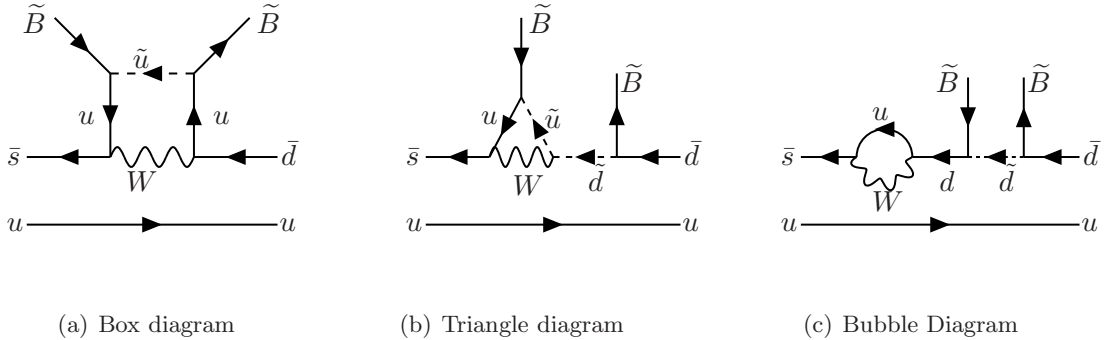
\begin{figure}
\centering
\subfigure[Box diagram\label{subfig:box}]{
\unitlength=1.5 pt
\SetScale{1.5}
\SetWidth{0.5}      
\begin{picture}(95,75)(0,0)
\Text(11,18)[l]{$u$}
\Text(84,18)[r]{$u$}
\ArrowLine(16,18)(79,18)
\ArrowLine(79,35)(58,35)
\ArrowLine(58,35)(58,56)
\ArrowLine(58,56)(72,70)
\ArrowLine(23,70)(37,56)
\ArrowLine(37,56)(37,35)
\ArrowLine(37,35)(16,35)
\Photon(37,35)(58,35){2.5}{3}
\DashArrowLine(58,56)(37,56){2}
\Text(11,35)[l]{$\bar{s}$}
\Text(84,36)[r]{$\bar{d}$}
\Text(49,29)[r]{$W$}
\Text(29,45)[l]{$u$}
\Text(63,45)[l]{$u$}
\Text(47,60)[r]{$\tilde{u}$}
\Text(15,70)[l]{$\bino$}
\Text(80,70)[r]{$\bino$}
\end{picture} \
}
\hspace*{-15mm}
\subfigure[Triangle diagram\label{subfig:tri}]{
\unitlength=1.5 pt
\SetScale{1.5}
\SetWidth{0.5}      
{} \qquad\allowbreak
\begin{picture}(95,75)(0,0)
\Text(11,18)[l]{$u$}
\Text(85,18)[r]{$u$}
\Text(11,35)[l]{$\bar{s}$}
\Text(85,36)[r]{$\bar{d}$}
\ArrowLine(16,18)(80,18)
\ArrowLine(64,35)(64,56)
\ArrowLine(80,35)(64,35)
\DashArrowLine(64,35)(48,35){2}
\Photon(48,35)(32,35){2.5}{3}
\ArrowLine(32,35)(16,35)
\ArrowLine(40,70)(40,50)
\ArrowLine(40,50)(32,35)
\DashArrowLine(48,35)(40,50){2}
\Text(38,28)[l]{$W$}
\Text(57,30)[l]{$\tilde{d}$}
\Text(30,45)[l]{$u$}
\Text(46,45)[l]{$\tilde{u}$}
\Text(42,70)[l]{$\bino$}
\Text(65,55)[l]{$\bino$}
\end{picture} \
}
\hspace*{-15mm}
\subfigure[Bubble Diagram\label{subfig:bubble}]{
\unitlength=1.5 pt
\SetScale{1.5}
\SetWidth{0.5}      
{} \qquad\allowbreak
\begin{picture}(95,75)(0,0)
\Text(5,18)[l]{$u$}
\Text(85,18)[r]{$u$}
\Text(5,35)[l]{$\bar{s}$}
\Text(85,36)[r]{$\bar{d}$}
\ArrowLine(10,18)(80,18)
\ArrowLine(80,35)(66,35)
\ArrowLine(66,35)(66,55)
\DashArrowLine(66,35)(52,35){1}
\ArrowLine(52,55)(52,35)
\ArrowLine(52,35)(38,35)
\ArrowArc(31,35)(7,0,180)
\PhotonArc(31,35)(7,180,0){1.5}{4}
\ArrowLine(24,35)(10,35)
\Text(73,55)[r]{$\bino$}
\Text(50,55)[r]{$\bino$}
\Text(60,30)[l]{$\tilde{d}$}
\Text(45,29)[l]{$d$}
\Text(28,24)[l]{$W$}
\Text(28,46)[l]{$u$}
\end{picture} \
}
\caption{Examples for loop induced supersymmetric Kaon decays}
\label{}
\end{figure}
In
Refs.~\cite{Gaillard:1982rw,Ellis:1982ve,Keung:1983hx,Nieves:1985ir},
the related photino decay $K^+\to\pi^+\photino\photino$ was
analysed. It can proceed at tree-level via the cascade decay
\begin{eqnarray}
K^+&\to&\pi^+\pi^0 \\
&& \quad\;\; \hookrightarrow \photino\photino \nonumber
\end{eqnarray}
It can also proceed directly at the loop level, for which example
Feynman graphs are shown in
Figs.~\ref{subfig:box}--\ref{subfig:bubble}.  It was found that for
large parts of the MSSM parameter space this decay is suppressed
relative to the SM decay $K^+\to\pi^+\nu\bar{\nu}$.  Since the SM
event rate is barely observable with present experiments, no bound on
the photino mass or the relevant sfermion masses in the propagators is
obtained. However, these results were derived when the correct mass of
the top quark was unknown.  Furthermore, only an incomplete set of the
relevant one-loop Feynman graphs was evaluated, and the neutralino was
restricted to be a photino.  Nonetheless, the estimate obtained using
the published results is well below the experimental bound, and we do
not expect any bound from present data for a bino. However, in
Ref.~\cite{work-ip:2008} this issue will be investigated based on a
complete analysis.

A similar analysis can be carried out for the decay $B^+
\to\pi\bino\bino$. The corresponding SUSY box diagrams involve $W^\pm$
bosons, squarks, charginos, and Higgs bosons.  As evaluated in
Ref.~\cite{Gaillard:1982rw,Ellis:1982ve}, the suppression by the
scalar fermions in the loops (which are assumed to exceed the
experimental bounds of $\sim 300 \GeV$) leads to a contribution that
is too small to set any limits on a light bino.

Another rare decay in which a light bino could in principle
play a role is $b \to s \gamma$. 
However, at the one-loop level neutralinos do not
contribute. The effect of a light $\bino$ at the two-loop level would
be well within the current theoretical and experimental
uncertainties~\cite{Isidori2008}.

}
\section{Astrophysical and Cosmological Bounds on the Neutralino Mass}
\label{ch:cosmo}
{
\input paperdef
\renewcommand{\bino}{{\widetilde B}} 
\newcommand{\dif}{\mathrm{d}} 

In this section we briefly consider the implications of a very light
neutralino for astrophysics and cosmology. We focus on supernova
cooling and the dark matter of the universe. In
Ref.~\cite{Jedamzik:2004ip} the implications for a moderately light
neutralino with $\mneu{1}\gsim5\,$ GeV for Big Bang nucleosynthesis
are discussed, which we shall not further consider here.

\subsection{Supernova Cooling}

During a supernova explosion neutrinos are abundantly produced in the
dense core. They \textit{diffuse} out with a time scale of $\mathcal{O} 
(10\,\mathrm{sec})$ \cite{Freedman:1973yd,Mazurek:1976kv,Sato:1975vu}.
The neutrinos with this time structure have indeed been observed after
Supernova 1987a \cite{Hirata:1987hu,Bionta:1987qt}. If light
neutralinos exist with mass less or of order the supernova core
temperature, $T_c={\cal O}(30\MeV)$, they can also be produced
abundantly during core collapse. Depending on their interactions,
these neutralinos escape freely from the supernova, rapidly cooling
the core \cite{Gandhi:1990bq}. As the temperature drops, the neutrino
scattering cross section drops with the square of the temperature,
leading eventually to free-streaming neutrinos.  Thus rapid cooling of
the supernova with a time scale well below 10 sec is excluded by the
neutrino observation from Supernova 1987a.  This can be used
to set restrictions on the light neutralino mass, as well as its
interactions. This was originally addressed for photinos in
Refs.~\cite{Grifols:1988fw,Ellis:1988aa,Lau:1993vf}. The authors of
Ref.~\cite{Dreiner:2003wh} derived important lower bounds on a light
bino, which we briefly summarize here; see also
Ref.~\cite{Kachelriess:2000dz}. Our focus here is on a massless
neutralino.

The two main neutralino production mechanisms in a supernova are
electron-positron annihilation and nucleon-nucleon ($NN$)
``neutralino-strahlung'':
\begin{eqnarray}
e^+ + e^-&\longrightarrow & \neu{1} + \neu{1}\,, \label{ul:Xs-e}\\
N+N&\longrightarrow& N+N+\neu{1}+\neu{1}\,. \label{ul:Xs-n}
\end{eqnarray}
Once produced, the neutralinos have a mean-free-path, $\lam_{\neu{1}}$,
in the supernova core which is determined via the cross sections for 
the processes
\begin{eqnarray}
\neu{1} +e &\longrightarrow& \neu{1} + e\,,\label{ul:scatter1} \\
\neu{1} + N &\longrightarrow& \neu{1} + N\,,
\label{ul:scatter2}
\end{eqnarray}
as well as the electron and nucleon densities.  If $\lam_{\neu{1}}$ is
of order of the core size, $R_c={\cal O}(10\,{\rm km})$, or larger, the
neutralinos escape freely and thus cool the supernova rapidly.
However, if the neutralinos have masses $\mneu{1}$ much greater than
the supernova core temperature $T_c$, then their production is
Boltzmann-factor suppressed and they affect the cooling negligibly,
independent of $\lam_{\neu{1}}$. Demanding that $\mneu{1}$ be large
enough that neutralino-cooling does not markedly alter the neutrino
signal---particularly its time-structure---allows one to set a lower
limit on the neutralino mass.  Note that this limit depends strongly
on the squark and selectron masses, which enter to the fourth power in
Eqs.~(\ref{ul:Xs-e}), (\ref{ul:Xs-n}) and Eqs.~(\ref{ul:scatter1}),
(\ref{ul:scatter2}) through the relevant propagators.

A proper treatment of this problem would be to expand the existing
supernova code(s) to include the production and the scattering of
neutralinos. Thus the neutralinos would be involved in the complete
time evolution of the supernova, which could affect the particle
densities within the supernova and the supernova temperature as a
function of time. Such a treatment is beyond the scope of this
paper. A good estimate of the effect of the neutralinos on the
supernova evolution can be obtained, if we use the existing codes
\cite{Burrows:1986me,Pons:1998mm,Mezzacappa:2000jb,Buras:2005rp} and
treat this non-supersymmetric supernova as a fixed
background,~\textit{i.e.} we assume the neutralino effect on the
evolution to be small. Using the resulting electron and nucleon
densities, we can compute the production and scattering of the
light neutralinos.  We then employ the Raffelt criterion
\cite{Raffelt:1996wa}, requiring that the maximal emitted energy from
the supernova via neutralino radiation is $\le 10^{52}\erg$. In
Ref.~\cite{Dreiner:2003wh}, it was then found that for selectron
masses in the range $300 \GeV \lsim m_{\tilde e}\lsim 900 \GeV$
neutralino masses below $100\MeV$ are excluded. As the selectron mass
is increased from 900 GeV the lower bound an the neutralino mass
gradually decreases. For selectron masses above $1.2\tev$ there is no
lower bound on the lightest neutralino mass.

Similar, however much less restrictive arguments also hold for the
squark mass dependence. For a massless neutralino, squark masses
between 300 GeV and 360 GeV are excluded.

For selectron and squark masses below 300 GeV the mean-free-path of
the neutralino is smaller than the supernova core size: $\lam_{
\neu{1}}<R_c$, \textit{i.e.} the neutralinos are trapped, and diffuse
out, just like the neutrinos. In this case the above approximate
procedure is no longer valid and the neutralinos must be included in
the numerical supernova simulation. This has to-date not been
performed. Thus at present, \textit{massless} neutralinos are not
excluded by the Supernova 1987a observations for $m_{\tilde
e},\,m_{\tilde q} < 300$~GeV or both $m_{\tilde e}> 1200$ GeV and
$m_{\tilde q}>360\,$GeV.

It should be pointed out that to-date it is not yet possible to
successfully simulate a full supernova explosion, in particular the
outgoing shock wave still stalls. Thus some ingredient is still
missing. More recently the simulations are being extended to three
dimensions with the inclusion of turbulent effects
\cite{Scheck:2003rw,Janka:2007di}. An eventual full solution could 
in principle lead to a modification of the above results.

\subsection{Dark Matter}

In the MSSM with conserved proton hexality or conserved $R$-parity,
the lightest neutralino is stable and will contribute to the dark
matter in the universe
\cite{Ellis:1983ew,Griest:1988ma,Jungman:1995df,Drees:1996pk,Ellis:2003cw}.
If it is very light, \textit{i.e.} relativistic at freeze-out, it will
contribute hot dark matter. If the dark matter candidate is
non-relativistic at freeze-out, it contributes to the cold dark matter
of the universe.  The structure formation in the early universe is
best described by cold dark matter alone \cite{Blumenthal:1984bp}.
Thus the contribution of hot dark matter to the energy density of the
universe is severely restricted. We discuss the resulting bound on the
light neutralino mass, the Cowsik--McClelland bound, in
Sect.~\ref{cowsik-mcc}.

For cold dark matter, it is well known that for smaller interaction
cross sections the resulting relic density is larger. Furthermore the
reaction rate also decreases with decreasing mass of the dark matter
candidate. Therefore, if we assumes that the candidate particle
provides the required cold dark matter of the universe, we obtain a
lower bound on the particle mass, the Lee--Weinberg bound. We shall
discuss the resulting bound for a neutralino in
Sect.~\ref{lee-weinberg}. This bound assumes the standard big-bang
cosmology with a radiation dominated universe prior to
nucleosynthesis. If one drops this assumption, it has recently
been shown that the lower bound is substantially weakened
\cite{Profumo:2008yg}. If for example one generalizes the MSSM to the
next-to-minimal supersymmetric Standard Model (NMSSM), which contains
an extra singlet chiral superfield, the bounds can also be weakened
\cite{McElrath:2005bp,Barger:2005hb,Gunion:2005rw}. We restrict
ourselves to the MSSM and the standard big-bang cosmology.

\subsubsection{The Cowsik--McClelland Bound}
\label{cowsik-mcc}

Here, we consider the case of a (nearly) massless neutralino,
$\mneu{1}\lesssim\mathcal{O}(1\, \mathrm{eV})$.  As for meson decays
above, we restrict ourselves to a pure bino neutralino
\cite{Choudhury:1999tn}. Since the very light bino contributes to the
hot dark matter of the universe, we assume here implicitly that the
cold dark matter originates from another source, see \textit{e.g.}
Ref.~\cite{Steffen:2008qp} for a review of alternate candidates. For a
sufficiently light bino the contribution of the hot dark matter to the
energy density of the universe is expected to be consistent with
present observations. We wish here to quantify this statement and thus
determine an upper mass bound on a very light stable bino.

Note that if we were to assume that proton hexality or $R$-parity is
violated, then for such a light bino [$\mneu{1}\lesssim \mathcal{O}
(1\, \mathrm{eV})$], kinematically the only open decay mode is
$\bino\rightarrow\nu\gamma$ (provided a lighter neutrino exists),
which proceeds via a one-loop diagram. Using the computation in
Ref.~\cite{Dawson:1985vr}, we can estimate the lifetime in this case as
\begin{equation}
\tau_\bino \approx \frac{1}{\lam^2} 10^{-9}
\,\mathrm{sec}\, \left(\frac{m_{\tilde f}}{100\,\GeV}\right)^4
\left(\frac{1\,\GeV}{m_\bino}\right)^3\,,
\end{equation}
where $\lam$ is the relevant $R$-parity violating coupling
\cite{b3msugra}, and $m_{\tilde f}$ is the mass of the sfermion
entering the loop. For $\lam=0.01$ (a typical upper bound), $m_{\tilde
  f}=100\GeV$ and a bino mass $m_\bino=1\,\mathrm{eV}$, we obtain a
lifetime of about $10^{22}\,$sec, well above the age of the universe.

For a stable bino, the bino relic energy density, $\rho_\bino$,
divided by the critical energy density of the universe, $\rho_c$, is
given by \cite{Kolb:1990eu}
\begin{eqnarray}
\label{ul:eq:relicdensity}
\Omega_\bino &\equiv& \frac{\rho_\bino}{\rho_c}
\;=\;  \frac{43}{11}\,\frac{\zeta(3)}{\pi^2}\,\frac{8\pi G_N}
{3 H_0^2}\,\frac{g_{\mathrm{eff}}(\bino)}{g_{\ast S}(T)}\,
            T_\gamma^3 \,m_\bino\,.
\end{eqnarray}
$G_N$ and $H_0$ denote Newton's gravitation constant and the present
value of the Hubble constant, respectively. $\zeta(3)$ is the Riemann
zeta function evaluated at 3.  $T_\gamma$ is the photon temperature.
Recall the \textit{effective} internal degrees of freedom of a
particle are given in terms of the internal degrees of freedom, $g$,
by
\begin{eqnarray}
g_{\mathrm{eff}} &=& \left\{
        \begin{array}{ll}
        \phantom{\frac{3}{4}}g, &\text{for a boson}\\[2mm]
        \frac{3}{4}g,& \text{for a fermion}
        \end{array}
        \right. \enspace .
\end{eqnarray}
For the bino, which is a Majorana fermion, we have:
$g_{\mathrm{eff}}(\bino)=2\times (3/4)=1.5$.  $g_{\ast S}$ is given by
\begin{eqnarray}
g_{\ast S}&=&\sum_{i=\text{bosons}}g_i\cdot\left(\frac{T_i}{T}
\right)^3+\frac{7}{8}\sum_{i=\text{fermions}}g_i\cdot\left(\frac{T_i}
{T}\right)^3\enspace.
\end{eqnarray}
Here, $g_i$ is the number of internal degrees of freedom and $T_i$ the
temperature of the particle species $i$, respectively. The sum runs
over the index $i$ for all species in thermal equilibrium at
temperature $T$.

In order for the bino hot dark matter not to disturb the structure
formation, we conservatively assume its contribution to the total
energy density of the universe to be less than the upper bound on the
energy density of the neutrinos, as determined by the WMAP data
\cite{Spergel:2006hy,Dunkley:2008ie}
\begin{equation}
\label{eq:bound}
\Omega_\bino h^2 \le  [\Omega_{\nu} h^2]_{\text{max}} = 0.0076\,.
\end{equation}

Light binos decouple at $T \approx \mathcal{O}(1-10\enspace
\mathrm{MeV})$. This temperature is somewhat higher than the
temperature where the neutrinos decouple, because the selectron mass
is larger than $\MW$. The higher selectron mass leads to a smaller
bino scattering cross section, \textit{i.e.} earlier freeze-out.
However, the temperature is well below the muon mass, so it is not
necessary to know the exact value.  Nevertheless at this temperature,
we have 2 bosonic and 12 fermionic relativistic degrees of freedom
(one photon, one Dirac electron, three left handed neutrino species,
and one light Majorana neutralino) leading to $g_{\ast S} = 12.5$.
{}From Eqs.~(\ref{ul:eq:relicdensity}) and (\ref{eq:bound}), we 
find the conservative upper bound
\begin{eqnarray}
m_\bino  \le 0.7 \enspace\mathrm{eV}\,.
\end{eqnarray}
Thus a very light bino with mass below about 1 eV is consistent
with structure formation. This line of argument was originally used by
Gershtein and Zel'dovich \cite{Gershtein:1966gg} and Cowsik and
McClelland~\cite{Cowsik:1972gh} to derive a neutrino upper mass bound,
by requiring $\Omega_\nu\leq1$. We have here obtained an upper
mass bound for a hot dark matter bino.

\subsubsection{The Lee--Weinberg bound}
\label{lee-weinberg}

In this subsection, we determine the \textit{lower} mass bound
on a light neutralino from the requirement that it alone
provides the required cold dark matter in the universe. This is based
on the original work for massive neutrinos
\cite{Lee:1977ua,Hut:1977zn,Sato:1977ye,Vysotsky:1977pe} and the
resulting bound is referred to as the Lee--Weinberg bound in the
literature. As a light neutralino, we consider here a neutralino with
a mass below the LEP bound, Eq.(\ref{lep-bound}), but which is 
nevertheless
non-relativistic in the early universe, \textit{i.e.}  neutralinos
with $\mneu{1} \geq \mathcal{O}(5\GeV)$. For simplicity, we shall
again restrict ourselves to the case of a pure bino, as we are mainly
interested in qualitative statements. This neutralino mass range has
been investigated numerically in
Refs.~\cite{Hooper:2002nq,Belanger:2002nr,Bottino:2002ry,Belanger:2003wb,Bottino:2003iu,Bottino:2003cz,Lee:2007ai}
in some detail. We present here a semi-analytical treatment of the
Boltzmann equation following Ref.~\cite{Kolb:1990eu}. This gives some
insight into the dependence of the neutralino relic density on the
main parameters in our case: the bino mass $m_\bino$, and the slepton
mass $m_{\tilde\ell}$. At the end of this section, we compare our
results to
Refs.~\cite{Hooper:2002nq,Belanger:2002nr,Bottino:2002ry,Belanger:2003wb,Bottino:2003iu,Lee:2007ai}.

For simplicity, we shall consider only the neutralino annihilation
into leptons
\begin{eqnarray}
\bino \bino \rightarrow \ell \overline{\ell},
    \quad \ell = e,\mu,\tau,\nu_e,\nu_\mu,\nu_\tau\,.
\end{eqnarray}
The $\tau$--lepton is considered as massless, all sleptons are assumed
to have common mass $m_{\tilde{\ell}}$. The relevant annihilation
cross sections are then related by
\begin{eqnarray}
\sigma(\bino\bino \rightarrow \ell^-_R \overline{\ell}^+_L)
= 16\, \sigma(\bino \bino \rightarrow \ell^-_L \overline{\ell}^+_R)
= 16\, \sigma(\bino \bino\rightarrow \nu_\ell \overline{\nu}_\ell)\,.
\end{eqnarray}
The cross section averaged over the thermal distributions of the
incoming particles is given by \cite{Kolb:1990eu,Gondolo:1990dk}
\begin{eqnarray}
\langle\sigma(\bino\bino \rightarrow \ell
  \overline{\ell})|v|\rangle \approx \sigma_0 x^{-1}\equiv
  54\pi\frac{\alpha^2}{\cw^4}
      \frac{m_{\bino}^2}{m^4_{\tilde{\ell}}}\,x^{-1}\,,\qquad 
\text{with }\,\;x\equiv\frac{m_\bino}{T}\,.
\label{xsec}
\end{eqnarray}
Here $T$ is the temperature of the universe.  Implementing our
specific cross section into the procedure outlined in Chapter 5.2 of
Ref.~\cite{Kolb:1990eu}, we obtain for the bino freeze-out temperature, $T_f$,
\begin{eqnarray}
x_f\,\equiv\,\frac{m_\bino}{T_f} &\approx & \ln\left[ \frac{0.19\cdot 
g}{g_{*}^{1/2}}
\left(\frac{{x \langle\sigma|v|\rangle s}}{{H(m_\bino)}}\right)_{x=1}\right] -
\frac{3}{2} \ln\left[\ln\left(
          \frac{0.19\cdot g}{g_{*}^{1/2}} 
          \left(\frac{x \langle\sigma|v|\rangle s}{H(m_\bino)}\right)_{x=1}
         \right)\right]\,.
\end{eqnarray}
Here the thermally averaged cross section is evaluated at $x=1$,
\textit{i.e.} $T=m_\bino$. The effective number of relativistic
degrees of freedom $g_\ast$ is given by
\begin{eqnarray}
\label{eq:gstar}
g_{\ast} &=& \sum_{i=\text{bosons}}g_i\left(\frac{T_i}{T}\right)^4 +
        \frac{7}{8}\sum_{i=\text{fermions}}g_i\left(\frac{T_i}{T}
\right)^4\,.\enspace
\end{eqnarray}
$H(m_\bino)$ is the Hubble parameter for $T=m_\bino$. The bino
contribution to the normalized energy density of the universe is then
\begin{figure}
\vspace*{8mm}
\centering
\scalebox{0.55}{\includegraphics{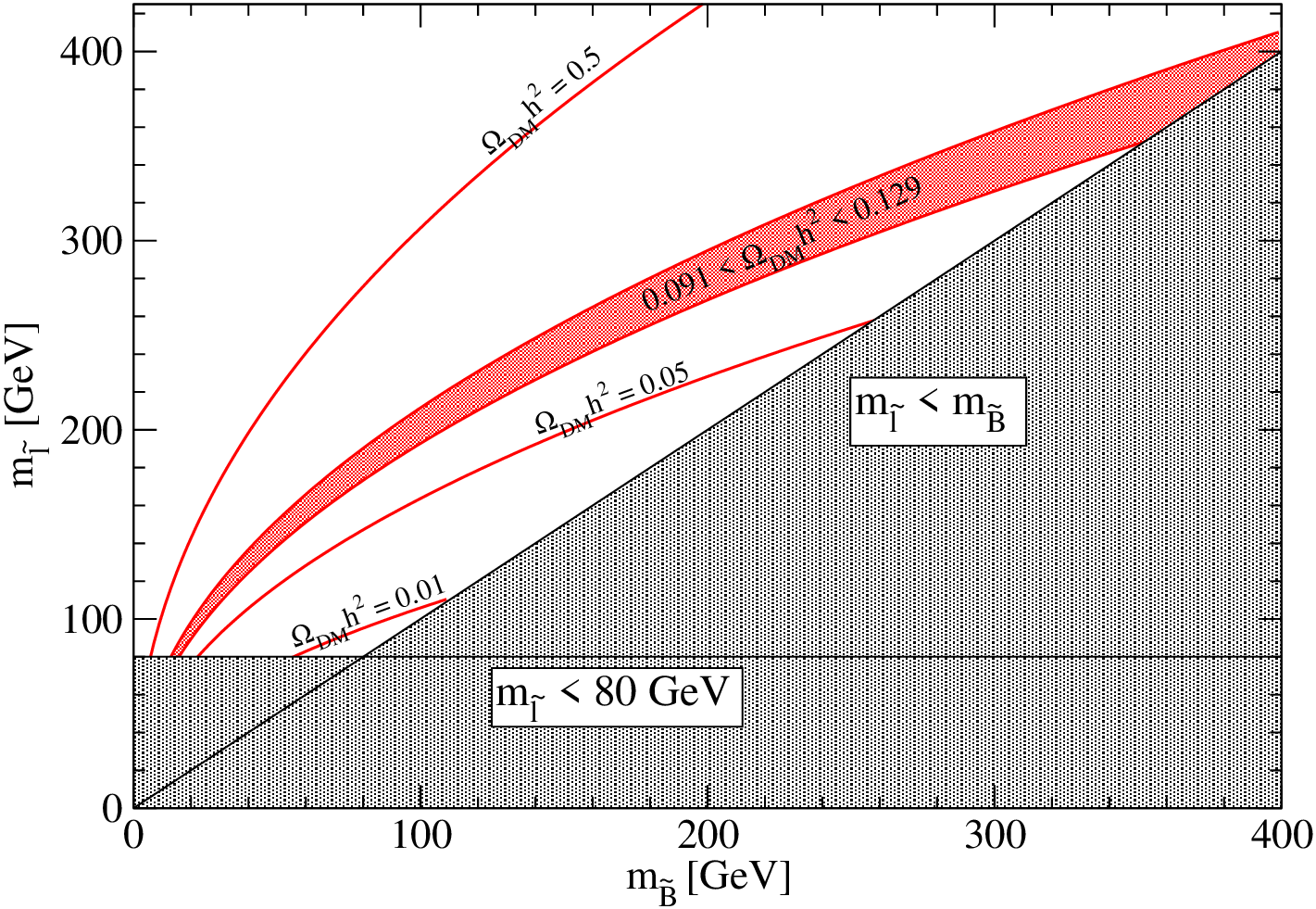}}
\caption{Constant contours of the relic density of a bino type LSP as
  a function of the LSP mass, $m_{\bino}$, and the slepton mass,
  $m_{\tilde\ell}$. In the red (dark) shaded area the relic 
  density is in
  the allowed range $0.091 < \Omega_{DM}h^2 < 0.129$. The grey
  (light) shaded areas are excluded due to LEP searches, $m_{\tilde{\ell}}
  <80\GeV$ and the theoretical requirement: $m_{\tilde\ell}<m_\bino$.}
\label{ul:fig:relic} 
\end{figure}
given by
\begin{eqnarray}
\label{ul:eq:reliclate}
\Omega_\bino h^2 = m_{\bino} n_\bino \approx
     \frac{2.14\times10^{9}x_f^{2}}{(g_{*S}/g_{*}^{1/2})
m_{\mathrm{Pl}}\sigma_0}
  \enspace\GeV^{-1}\,.\enspace
\end{eqnarray}
Here $m_{\mathrm{Pl}}$ denotes the Planck mass.
We have plotted in Fig.~\ref{ul:fig:relic} contours of
constant relic density in the $m_{\bino}$--$m_{\tilde\ell}$-plane.
The lower right-hand triangle of the figure is excluded since here the
sleptons are lighter than the neutralino, contrary to our assumption.
The bottom horizontal band, $m_{\tilde \ell}<80\,$GeV, is excluded due
to LEP searches \cite{Amsler:2008zz}.  The red (dark shaded) band
denotes the preferred relic density region from the 5-year WMAP data
\cite{Dunkley:2008ie}
\begin{equation}
\Omega_\bino h^2 = 0.1099\pm 3\sigma_\Omega = 0.1099\pm 0.0186\,,
\end{equation}
where we have implemented the 3 sigma bounds, and where the 1 sigma
error is $\sigma_\Omega=0.0062$.

As expected, we see in Eq.~(\ref{ul:eq:reliclate}) that the resulting
relic density is inversely proportional to the bino annihilation
cross section. Furthermore
\begin{equation}
\Omega_\bino h^2 \,\propto\, \frac{m_{\tilde\ell}^4}{m_\bino^2}\,.
\end{equation}
Thus for fixed $\Omega_\bino h^2$ the slepton mass $m_{\tilde\ell}
\propto \sqrt{m_\bino}$. This can be seen in
Fig.~\ref{ul:fig:relic}. Since observationally there exists an upper
bound $\Omega_\bino h^2<[\Omega_\bino h^2]_{\text{max}}=0.129$, this
translates into a lower bound on the bino mass $m_\bino$. We obtain
the most conservative bound for the smallest allowed slepton mass
$m_{\tilde\ell}=80\,$GeV. Inserting these numbers we obtain
\begin{equation}
m_\bino>13\,\text{GeV}\,.
\end{equation}
From Fig.~\ref{ul:fig:relic} it should also be clear that one obtains
an upper bound on $m_\bino$ for $m_{\tilde\ell}=m_\bino$ and
$\Omega_\bino h^2=[\Omega_\bino h^2]_{\text{max}}$,
\begin{equation}
m_\bino<419\GeV\,.
\end{equation}
This is then also the upper bound on the slepton mass in this scenario.

The neutralino relic density has been considered extensively in
  the literature, see for example
  Refs.~\cite{Pagels:1981ke,Goldberg:1983nd,Ellis:1983ew,Griest:1988ma,Jungman:1995df,Drees:1996pk,Ellis:2003cw}.
  More recently the question has been raised precisely how light the
  lightest neutralino can be, while still providing the entire (cold)
  dark matter in the universe as required by the WMAP 
  data~\cite{Hooper:2002nq,Belanger:2002nr,Bottino:2002ry,Belanger:2003wb,Bottino:2003iu,Bottino:2003cz}.
  The authors allowed for non-universal gaugino masses, \textit{cf.}
  Eq.~(\ref{assump}), as well as a general neutralino admixture. They
  took into account all kinematically allowed annihilation products,
  as well as co-annihilation and resonant annihilation. In
  Ref.~\cite{Hooper:2002nq} a lower bound of about $18 \GeV$ was found
  for a non-relativistic neutralino. The lowest lower bound of about
  $6 \GeV$ was found in Ref.~\cite{Bottino:2003iu,Belanger:2003wb} in models
  with a light pseudoscalar Higgs $A$ with mass $M_A<200 \GeV$. In
  Refs.~\cite{Bottino:2002ry,Bottino:2003iu} a semi-analytical
  approximation was also performed. They focused on the $s$-channel
  Higgs exchange specifically in the low $M_A$ region. It is
  encouraging how well these numbers agree with our approximate
  results, which only considered the neutralino annihilation to
  leptons. This is mainly because the resonant and co-annihilation
  effects we have ignored are not relevant in the low-mass region.

In a further study, it was found that when allowing for explicit CP
violation in the Higgs, and thus the neutralino sector the lower
bound is reduced to 3 GeV \cite{Lee:2007ai}.

}

\section{Summary and Conclusion}
{
\input paperdef

In this paper we have studied mass bounds on light neutralinos from
collider physics, precision observables, meson decays, and
astrophysics and cosmology. We have focused on the question how light
the $\neu{1}$ can be, in particular whether an essentially massless
$\neu{1}$ is allowed by the experimental and observational
data. Assuming a semi-simple gauged Lie group the parameters $M_1$ and
$M_2$ are independent. A very light neutralino is possible by mildly
tuning $M_1$, resulting in a substantial bino component of the
$\neu{1}$.

Specifically, we have analysed the processes $e^+e^- \to
\neu{1}\neu{2}$ and $e^+e^- \to \neu{1}\neu{1}\gamma$ for a massless
neutralino. The LEP searches have determined an upper bound on the
cross section of the first process. This reaction depends on the
selectron masses, due to the $t$-channel selectron exchange
contribution. Consequently, we can translate the upper bound on the
$\neu{1}\neu{2}$ production cross section into a lower mass bound on
the selectron masses, depending on the choice of $\mu$ and
$M_2$. Assuming degenerate selectron masses and making the
conservative assumption that the branching ratio of $\neu{2}$ into 
$\neu{1}$ and a $Z$ boson is 100\%, for $\mu < 150\GeV$ we
find $m_{\tilde e} \gsim 1$~TeV.  For $\mu \gsim 200\GeV$ no bounds on
$m_{\tilde e}$ are obtained. As a consequence, 
this search channel yields no bounds on the mass of the lightest
neutralino. Radiative
neutralino production, $e^+e^- \to \neu{1}\neu{1}\gamma$, which has
been searched for at LEP2 and $b$-factories, also does not result in
any bound on $\mneu{1}$.

Electroweak precision observables such as the $W$~boson mass, $\MW$,
the effective leptonic weak mixing angle, $\sweff$, or the invisible
$Z$~width, $\Ga_{\rm inv}$ are potentially sensitive to a very light
neutralino. 
For $\MW$ and $\sweff$, where the neutralino contribution
is a loop effect, we find that even the most precise anticipated
measurements will not be able to constrain the mass of $\neu{1}$.
The contribution to the invisible $Z$~width, arising from the decay of
the $Z$ boson into a pair of very light neutralinos,
can give rise to a measurable effect if the bino component of the $Z$
boson is not too high. Also this constraint, however, is not very
powerful for most of the parameter space, and the largest effects occur
in parameter regions that are already excluded by the LEP searches
for charginos.
We have furthermore analyzed the effect of a light
neutralino on electric dipole moments and the anomalous magnetic
moment of the muon. Both do not show any relevant sensitivity
to $\mneu{1}$.

Next we investigated the bounds from rare meson decays.  Pseudo-scalar
mesons and vector mesons can decay into a pair of massless
neutralinos. This decay, however, is suppressed by (large) squark
masses and, in the case of pseudo-scalar mesons,
a small neutralino mass. This decay has the same signature
as the SM decay into a neutrino pair.  Using the corresponding
experimental upper bounds no constraints on $\mneu{1}$ could be
established. We also found no bound on $\mneu{1}$ from the
rare, loop-induced cascade decay $K^+\rightarrow\pi^+\neu{1}\neu{1}$.

We next considered the astrophysical bounds on the neutralino
mass from neutralino production in a supernova explosion.  We found
that a massless neutralino is consistent with the neutrino
observations of supernova 1987a, provided the selectron mass (which
enters in the production and scattering cross sections) is larger than
1.2 TeV and the squark mass is larger than 360 GeV. However, for
squark and selectron masses below 300 GeV the neutralinos are trapped
in the supernova, similar to the neutrinos. 
This scenario has not yet been sufficiently analyzed, so that
also in this case a massless neutralino cannot be ruled out at present.

Finally, we considered cosmological bounds on the neutralino mass. If
the neutralino constitutes hot dark matter it is in agreement with the
WMAP observations and structure formation for a mass \textit{below}
1 eV. For the case that the lightest neutralino constitutes the
entire cold dark matter of the universe, we performed a
semi-analytical analysis and found a \textit{lower} bound of
$m_{\neu{1}}>13\,$GeV. This is in remarkably good agreement with the
more complete numerical computations which find a lower bound of about
6 GeV. 

Overall, we have found that a massless neutralino is consistent
with all present laboratory experiments and astrophysical and
cosmological observations. At the upcoming experiments at the
LHC very light neutralinos could be produced in cascade decays of
heavier supersymmetric particles. The anticipated precision in the
determination of the differences between squared masses could allow to
further test scenarios with very light neutralinos. Further insights can
be expected from high-precision measurements at a future $e^+e^-$ Linear
Collider.

}

\section{Acknowledgements}
We would like to thank Sebastian Grab, Daniel Koschade, Michael
Kr\"amer, Greg Landsberg, Ben O'Leary, Peter Richardson and Subir
Sarkar for numerous discussions on various aspects of this paper. HD
would like to thank the Aspen center for Physics where some of this
work was performed. HD and UL would like to thank Christoph Hanhart
and Daniel Phillips for the earlier collaboration on light neutralino
supernova bounds. The work of HD, OK and UL was supported by SFB TR-33
The Dark Universe.  The work of SH was partially supported by CICYT
(grant FPA2006--02315).  Work supported in part by the European
Community's Marie-Curie Research Training Network under contract
MRTN-CT-2006-035505 `Tools and Precision Calculations for Physics
Discoveries at Colliders' (HEPTOOLS).
This work has been supported by MICINN project FPA.2006-05294.


\end{document}